\documentclass[]{interact}

\usepackage{epstopdf}% To incorporate .eps illustrations using PDFLaTeX, etc.
\usepackage[caption=false]{subfig}% Support for small, `sub' figures and tables

\usepackage[doublespacing]{setspace}% To produce a `double spaced' document if required
\usepackage[numbers,sort&compress]{natbib}% Citation support using natbib.sty
\usepackage{xcolor}
\usepackage{algorithm}
\usepackage{algpseudocode}
\usepackage[title]{appendix}
\usepackage{dsfont}
\usepackage{booktabs}
\usepackage{colortbl}
\usepackage{caption}

\usepackage{wrapfig}
\usepackage{lscape}
\usepackage{rotating}
\usepackage{epstopdf}

\bibpunct[, ]{[}{]}{,}{n}{,}{,}% Citation support using natbib.sty
\theoremstyle{plain}% Theorem-like structures provided by amsthm.sty

\theoremstyle{definition}

\theoremstyle{remark}

\definecolor{intnull}{RGB}{213,229,255}

\begin{document}

\articletype{MANUSCRIPT}% Specify the article type or omit as appropriate

\title{Modeling non-linear spectral domain dependence using copulas with applications to rat local field potentials}

\author{
\name{Charles Fontaine\textsuperscript{a}\thanks{CONTACT C.~F. Email: charles.fontaine@kaust.edu.sa} and Ron D. Frostig\textsuperscript{b} and Hernando Ombao\textsuperscript{a}}
\affil{\textsuperscript{a} Statistics Program, King Abdullah University of Science and Technology (KAUST), Thuwal, Saudi Arabia 
}
\affil{\textsuperscript{b} Department of Neurobiology and Behavior, University of California-Irvine, U.S.A. 
}}

\maketitle

\begin{abstract}
This paper intends to develop tools for characterizing non-linear spectral dependence between spontaneous brain signals. We use parametric copula models (both bivariate and vine models) applied on the magnitude of Fourier coefficients rather than using coherence. The motivation behind this work is an experiment on rats that studied the impact of stroke on the connectivity structure (dependence) between local field potentials recorded at various channels. We address the following major questions. First, we ask whether one can detect any changepoint in the regime of a brain channel for a given frequency band based on a difference between the cumulative distribution functions modeled for each epoch (small window of time). Our proposed approach is an iterative algorithm which compares each successive bivariate copulas on all the epochs range, using a bivariate Kolmogorov-Smirnov statistic. Second, we ask whether stroke can alter the dependence structure of brain signals; and examine whether changes in dependence are present only in some channels or generalized across channels. These questions are addressed by comparing Vine-copulas models fitted for each epoch. We provide the necessary framework and show the effectiveness of our methods through the results for the local field potential data analysis of a rat.
\end{abstract}

\newpage

%Introduction
\section{Introduction}

\begin{figure}
\begin{center}
\includegraphics[scale=0.60 ]{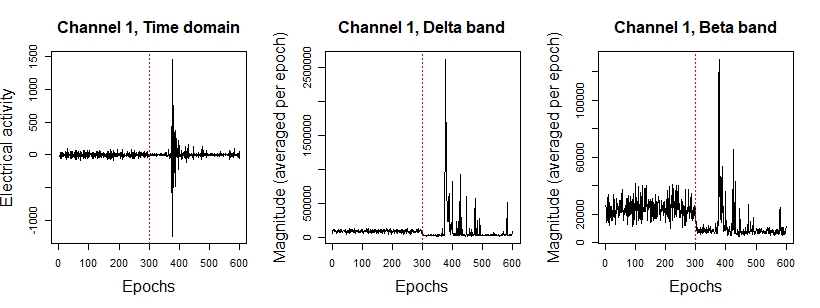}
\caption{Microelectrode (channel) $1$ for rat id $141020$. Red dotted line represents the onset of the induced stroke. \textbf{Left:} Local field potential recording on channel $1$. \textbf{Middle:} Plot of the magnitude of Fourier coefficients in the delta band across all epochs. \textbf{Right:} Plot of the magnitude of Fourier coefficients in the beta band across all epochs.}
\label{channel1Intro}
\end{center}
\end{figure}

Brain stroke occurs when blood circulation in one of the cerebral blood vessels is abnormally weak, and in such case, leads to death of the cells.. Brain stroke has been studied for years by biologists and neurologists. Studying this disorder  from the perspective of the changes in the brain’s electrical activities among different regions has yielded many clinically important results: these changes are so important that often they do irreversible damages to patients and incur extravagant costs to society (\textit{e.g., high medical expenses and low quality of patients’ lives}). In order to reduce these societal costs, neuroscientists study the behavior of the cortex activity by inducing stroke in rats. Due to ethical considerations, stroke experiments are conducted mostly only on rats. This paper is based on an experimental setup designed to induce stroke in a rat and to study the electrical oscillations among different regions in his brain.  Using the copula information, we developed methods for assessing and analyzing dependence between the rat’s brain regions. Our work is in collaboration with neuroscientists from University of California at Irvine (co-author Frostig and student \citet{wann2017}) who mechanically induced brain stroke in the rats by clamping a brain artery and recorded the brain activity on $32$ microelectrodes (or channels) before and after the stroke. Figure \ref{channel1Intro} shows, for one of the analyzed rats, how the data act differently in the pre-stroke phase (first $5$ minutes or first $300$ epochs) versus in the post-stroke phase (last $5$ minutes or last $300$ epochs). The detailed setup is described in Section \ref{changepoint}.

One goal is to analyze the changes in the dependence between some channels for all frequency bands by using flexible models. Most analyses use coherence or correlation which are simple to implement but they are severely limited because they capture only linear dependence structures. Thus, we present an innovative methodology based on the notion of copula function to capture the complexity of the dependence and by comparing two (or more) copulas. Moreover, we assess whether or not the dependence between pairs of epochs changes across ($1$-second) epochs of the entire recording period. From the recorded electrical activity during the laboratory experiment, the impact of brain stroke is observable for most microelectrodes on almost all frequency bands; and the effect of the stroke on brain signals appear to last throughout the entire post-stroke recording (see Appendix \ref{appChannels}). Our contributions in this paper are the following. First, we present an algorithm to help recognize which epoch(s) exhibit changes in the dependence structure of the brain signals. This recognition of a changepoint is key to understanding the biological mechanisms occurring in the time window between the onset of the stroke and the moment where significant changes occur. Second, we present a method to assess if the dependence structure during pre-stroke differs from the one from post-stroke. This assessment is crucial to understand if this channel will be impacted by the stroke. This method is also used later to compare the dependence structure among two different channels for a given frequency band. 

In the literature, many studies investigated changes in dependence for brain channels (in electroencephalograms) defined in the spectral domain. Among them, we highlight \citet{ombao2016}, \citet{fiecas2016}, \citet{long2004}, \citet{purdon2001}, \citet{nunez2006} and \citet{gotman1982}.  However, the primary limitation of these studies is that they look only into the linear dependence between signals. Thus, they could miss potential complex (or non-linear) dependence structure between signals.  Most methods reported explored the problem of detecting  one (or many) changepoint moment(s) (e.g., within an epoch). A major approach is based on segmentation of the series in order to assess a possible discrepancy between these segments: on either a change in mean or a difference in the correlation structure. Many authors considered the segmentation: e.g., \citet{adak1998} with binary trees and windowed spectra to adaptively partition data; \citet{ombao2001} derive a segmentation by selecting the best localized basis from the SLEX (smooth local exponential) library. Another example is the estimation of a penalized minimum contrast (\citet{lavielle1999}). Its  principle has two steps. In the first step, a contrast function is computed over a segments of a time period (or a sequence defined in the frequency domain - see \citet{lavielle2000}). The changepoints are then selected to be a solution to the minimization problem. Another example of that segmentation is based on probabilistic pruning methods. The principle of pruning is to predict the probability that a segment belongs to a stationary process rather than its likelihood. This method has been well studied by \citet{james2015} and \citet{kifer2004}. Another approach presented by \citet{davis2006} is the Auto-PARM: it consists in fitting multiple auto-regressive (AR) functions to segments of time. But fitting the AR model could be subject to model misspecification. The third kind of methods for detecting changepoints is based on hypothesis tests. \citet{dette2009} and \citet{dette2012} proposed an approach to test the equality of spectrum between two successive segments. This idea iis interesting but it does not take into account the nature and the structure of the dependence between these successive segments. 

The use of the joint cumulative distribution functions with brain signals has also been explored to study dependence between random variables in general. These functions, namely \textit{copula} models, have the main advantage to represent the dependence as functions that provide the information of both ''strength'' and ''structure'' of the relation between two variables. For example, in Figure \ref{channel1Intro}, for the three cases, it is obvious that the dependence pattern between succeeding epochs during pre-stroke (first $300$ epochs) and the one during post-stroke (last $300$ epochs) are different and that the dependence structure from epoch $300$ to epoch $400$ is not the same than the one between epoch $500$ and epoch $600$. These particularities in dependence structure will be fully detected with a copula under a right specification. \citet{iyengar2010} used it to quantify synchronicity between multiples electroencephalographic (EEG) signals. \citet{dauwels2012} used copulas in their attempt to design brain network. \citet{ince2017} presented a framework to assess dependence for neuroimaging data based on the gaussian copula. Even if all of these approaches presented a copula-based framework for brain signals data, none of them was interested in a detection of a change (or of a changepoint) between successive epochs.

\begin{figure}[!htbp]
\begin{center}
\includegraphics[scale=0.72 ]{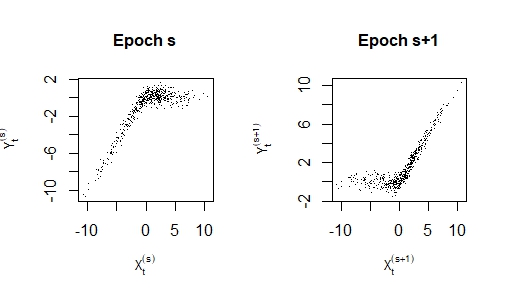}
\caption{Illustration of the example motivating the use of copulas. \textbf{Left:} Scatterplot of $X_t^{(s)}$ and $Y_t^{(s)}$. \textbf{Right:} Scatterplot of $X_t^{(s+1)}$ and $Y_t^{(s+1)}$. The correlation at epochs $s$ and $s+1$ are almost the same but it is clear that the dependence structure is different.}
\label{ExampleIntro}
\end{center}
\end{figure}

To show the advantage of assessing dependence through a copula function instead of via standard linear correlation-based methods, consider the following basic example. This example mimics the properties of rat local field potentials in this paper.For $t=1,...,500$, let $X_t^{(r)}$ and $Y_t^{(r)}$ be two random variables following the same dependence path for epochs $r=1,...,s$ such that $X_t \sim AR(1)$ of parameter $\phi=0.9$ and $Y_t^{(r)}=\mathcal{D}(X_t^{(r)})X_t^{(r)}+\epsilon_t^{(r)}$ where $\epsilon^{(r)}_t$ is a zero mean unit variance noise and $\mathcal{D}(X_t^{(r)})$ is the logistic curve $\exp\{-X_t^{(r)}\}/(1+\exp\{-X_t^{(r)}\})$. For epochs $r=s+1,...,R$, $X_t^{(r)}$ keeps following the same autoregressive process, but $Y_t^{(r)}=\mathcal{D}'(X_t^{(r)})X_t^{(r)}+\epsilon_t^{(r)}$ where $\mathcal{D}'(X_t^{(r)})=\exp\{X_t^{(r)}\}/(1+\exp\{X_t^{(r)}\})$. Thus, a changepoint in the dependence structure is present between epochs $s$ and $s+1$. Under this setup, a correlation-based changepoint detection method will not detect the change because the correlation between $X_t^{r}$ and $Y_t^{r}$ at epoch $r=s$ is not different from the correlation at epoch $r=s+1$. Theoretically, Pearson's correlation will stay equal to approximately $0.80$. On Figure \ref{ExampleIntro}, one observes that, for epochs $r=1,...,s$, dependence is high in the lower tail and small in the upper tail; and for epochs $r=s+1,...,R$, one observes exactly the converse. However, the copula function catches these changes in the dependence structure. Indeed, under a right specification, two different copula models will be fitted: one for epochs $r=1,...,s$ and a completely different one for epochs $r=s+1,...,R$. Thus, with an adequate methodology to assess the equivalence between two copulas as discussed in this paper, a copula-based method will detect the changepoint between epochs $s$ and $s+1$, for which a correlation-based method fails.

We present in this article a copula-based framework to analyze changes between brain signals on given frequency bands for three different contexts. Firstly, we are interested in the detection of one (or many) changepoint(s) in the regime of a brain channel for a given frequency band. Secondly, we compare, within a single channel (microelectrode), if there is a difference in the dependence between successive epochs across the pre-stroke and post-stroke epochs. Thirdly, we compare the dependence structure of two different microelectrodes still on a given frequency band, on the entire recording time of $5$ minutes priot to and $5$ minutes post to the induced stroke. 

The remainder of this paper is organized as follows. In Section \ref{notation}, we present present briefly the transformation of time data to spectral data as well as the copula function in order to introduce our notation. In Section \ref{section2}, we present the necessary theoretical background to introduce our models and algorithms. Then, in Sections \ref{changepoint}, \ref{prepost} and \ref{twoChan}, we present analyses of the local field potential data recording during a span of $10$ minutes ($5$ minutes pre-stroke and $5$ minutes post-stroke). Our methodology directly applied to these data shows its performance by, at first, assessing the statistically significant changepoints in dependence between successive epochs for some specific channels. Secondly, it shows that the whole dependence structure between pre-stroke epochs is not significantly different, for all the channels, than the one for post-stroke epochs.
\section{Statistical prologue and notation}\label{notation}

To facilitate ease of reading of the paper, we include the notations in Appendix \ref{appendixNotation}. Let ${\bf{X}}=\left[ X_1,X_2,...,X_d\right]$ a three-dimensional matrix of dimension $T \times d \times R$  ($d$ brain channels divided into $R$ possibly over-lapping epochs (i.e., equal segmentation of $1$ second into $T$ time points)). Thus, one represents a single element of that matrix by $X_\ell^{(r)}(t)$, $\ell=1,...,d$; $ t=1,...,T$; $ r=1,...,R$, which might be seen as any recorded measure on channel $\ell$ during epoch $r$. Therefore, the 3-dimensional matrix is composed of $R$ matrices of size $T \times d$ denoted by ${\bf X}^{(r)} = [X_1^{(r)}, \ldots, X_{d}^{(r)}]$.

This paper focuses on dependence among brain channels in the frequency domain. We remark that from the experimental perspective, the channels are defined by microelectrodes different parts of the brain. The Fourier coefficient for the channel $\ell=1,...,d$, at epoch $r=1,...,R$ and at fundamental Fourier frequency $\omega_k = k/T$ is defined to be

$$f_{\ell \omega_k}^{(r)} = \frac{1}{\sqrt{T}} \sum_{t=1}^{T} X_\ell^{(r)}(t) 
\exp(-i 2 \pi \omega_k t).$$ Because this transformation outputs single frequencies and in our context we are interested in frequency bands, we have to segregate frequencies according to the bands' ranges and to smooth their magnitudes by averaging.

In this paper, we will study the dependence of magnitudes of the Fourier coefficients (or square roots of periodograms) between the different pairs of channels $\ell$ and $ \ell'$; $\ell,\ell'=1,...,d$ for the same epoch $r$. In addition, we will investigate the dependence between successive pairs of epochs $r$ and $r+1$. We denote the frequency bands by $\Omega_\kappa$ where $\kappa=1,...,Q$ represents the frequency band, and where $Q$ is the number of frequency bands considered in this paper (see Section \ref{section2} for details about the chosen bands). Thus, $\Omega_\kappa$ is a collection of fundamental Fourier frequencies $\omega_j$, $j=\text{index}(\Omega_\kappa^{(\text{min})}),...,\text{index}(\Omega_\kappa^{(\text{max})})$ where $\text{index}(\Omega_\kappa^{(\text{min})})$ is the index of the first value constituting the frequency band among the $k$ fundamental frequencies, and $\text{index}(\Omega_\kappa^{(\text{max})})$ is the one of the last frequency constituting that band.

 We now define ${\bf{\delta}}^{(r)}_{\Omega_\kappa} = [\delta_{1,\Omega_\kappa}^{(r)},...,\delta_{d,\Omega_\kappa}^{(r)}]$ which is the matrix of dimension $\text{card}(\Omega_\kappa) \times d$ where any column is a different channel $\ell=1,...,d$. Therefore, each column is represented by $\delta_{\ell,\Omega_\kappa}^{(r)}=[| f_{\ell,\Omega_\kappa^{(\text{(min)})}}^{(r)}|,...,| f_{d,\Omega_\kappa^{(\text{(max)})}}^{(r)}|]'
$, a vector of length $\text{card}(\Omega_\kappa)$ containing the magnitude for each Fourier fundamental frequency constituting the frequency band $\Omega_\kappa$ at epoch $r$. Hence, in the rest of this paper, we will consider $\delta_{\ell, \Omega_\kappa}^{(r)}, \delta_{\ell', \Omega_\kappa}^{(r)}$ as the random vectors on which our methodology is applied.

\subsection*{Copula function}

Let the brain channels be indexed by $\ell,\ell' \in \{1,2,...,d\}$, let the epochs be indexed by $r=1,...,R$ and let the frequency bands of interest to be $\Omega_{\kappa}$ and $\Omega_{\kappa'}$. Our goal is to assess the dependence between $\delta_{\ell,\Omega_\kappa}^{(r)},\delta_{\ell ',\Omega_{\kappa '}}^{(r')}$ in the cases where $\ell=\ell', r\neq r', \Omega_\kappa=\Omega_{\kappa'}$, where $\ell \neq\ell', r=r', \Omega_\kappa=\Omega_{\kappa'}$ and where $\ell =\ell', r=r', \Omega_\kappa \neq \Omega_{\kappa'}$, we will express the dependence between these two quantities by expressing their joint cumulative distribution function. To this end, one writes $H_{(\ell,\Omega_\kappa)(\ell ', \Omega_{\kappa '})}^{(r,r')}(\delta_{\ell,\Omega_\kappa}^{(r)},\delta_{\ell ',\Omega_{\kappa '}}^{(r')})$ as the bivariate joint distribution for the random variables $\delta_{\ell,\Omega_\kappa}^{(r)}$ and $\delta_{\ell ',\Omega_{\kappa '}}^{(r')}$. We denote its marginal distributions by $H_{\ell,\Omega_\kappa}^{(r)}(\delta_{\ell,\Omega_\kappa}^{(r)})$ and $H_{\ell',\Omega_{\kappa'}}^{(r')}(\delta_{\ell ',\Omega_{\kappa'}}^{(r')})$. Using Sklar's theorem\cite*{sklar1959}, this joint distribution can be rewritten in terms of a unique copula: 
$$
H_{(\ell,\Omega_\kappa)(\ell ', \Omega_{\kappa '})}^{(r,r')}(\delta_{\ell,\Omega_\kappa}^{(r)},\delta_{\ell ',\Omega_{\kappa '}}^{(r')})=C_{(\ell,\Omega_\kappa)(\ell ', \Omega_{\kappa '})}^{(r,r')}\left( H_{\ell,\Omega_\kappa}^{(r)}(\delta_{\ell,\Omega_\kappa}^{(r)}),H_{\ell',\Omega_{\kappa'}}^{(r')} (\delta_{\ell ',\Omega_{\kappa '}}^{(r')}) \right)
 $$ where $C$ is the exact copula linking $\delta_{\ell,\Omega_\kappa}^{(r)}$ to $\delta_{\ell',\Omega_{\kappa'}}^{(r')}$. \citet{fontaine2018} provided an inferential framework for such a joint model in the spectral domain. For the rest of this paper, in the case of $\ell=\ell'$ and $\kappa=\kappa'$, we reduce this notation to $C_{\ell,\Omega_\kappa}^{(r,r')}$. We also assume that the copulas are fully parametric meaning that either the copula structures than the marginal distributions are parametric. Furthermore, we assume the true copula parameter $\breve{\theta}$ to be inferred in two possible ways (depending on the clinical question we are trying to answer): by a maximum likelihood estimation denoted $\hat{\theta}_{{\bf K}}$ or by the inversion of Kendall's tau method, denoted $\hat{\theta}_\tau$. We remark that although many parametric families of copulas have been studied in the literature (see \citet{genest1986} or \citet{nelsen2007}), selecting a suitable copula model may be tricky. Therefore, in Section \ref{copulaSel}, we discuss the selection of a copula model and the impact of misspecification.

%Section presenting LFP data and methodology common to the methods:
\section{Theoretical framework related to copulas and distributions}\label{section2}

Prior to any statistical modeling, we applied a Fourier transform to the multichannel multiple-epoch brain signals, considering each microelectrode $\ell$ at each epoch $r$ as a single data vector. This transform was in order to obtain the values of the periodograms and then compute the magnitudes of these Fourier coefficients. As mentioned earlier, we decided to use frequency bands rather than single frequencies. To determine the range of the considered bands, we based our choice on a classical text in the literature: \citet{buzsaki2006} about the rhythms of the brain. Hence, we adopted the following bands: $\Delta \in (0,4)$Hertz, $\theta \in (4,8)$Hertz, $\alpha \in (8,12)$Hertz, $\beta \in (12,30)$Hertz and $\gamma \geq 30$ Hertz. We note that in our applications, we truncated $\gamma$ at $300$ Hertz and applied a notch filter to remove the $60$ Hertz activity.

Before describing the specific cases where we assess the dependence among brain channels for particular frequency bands, we first discuss the application of the Kolmogorov-Smirnov statistic to the multivariate setting. Kolmogorov-Smirnov might be used to compare two cdf together. In the univariate case, if $A(x)$ and $B(x)$ are two cdf, to test $H_0: A=B$ versus $H_1:A \neq B$, we use the statistic $$ D = \sup_{x \in \mathbb{R}} \|A(x)-B(x) \|,$$ which is known to converge almost surely to 0 under $H_0$ due to Donsker's theorem\cite*{donsker1952}.

In this paper, we are interested in the empirical value of that statistic in a multivariate context. We use it in different ways according what we study. However, the way that we deal with the statistics of test remains the same. In a bivariate situation, let ${\bf{X}}=(X_1,X_2)'$ and ${\bf{Y}}=(Y_1,Y_2)'$ be two random variables with respective joint cdf $A$ and $B$. Also, let $u,v$ be two finite partitions in any closed subset of $\mathbb{R}^2$, large enough to contain the supports of $\bf{X}$ and $\bf{Y}$. Hence, we define our computational approach of the bivariate Kolmogorov-Smirnov statistic as
\begin{eqnarray*}
D(u,v)&=& \sup_{(u,v)} \left| A(u,v)-B(u,v)  \right| \\
&=& \sup_{(u,v)} \left| C_{\bf{X}}(A_{X_1}(u),A_{X_2}(v))-C_{\bf{Y}}(B_{Y_1}(u),B_{Y_2}(v))  \right|
\end{eqnarray*}  
where $C_{\bf{X}},C_{\bf{Y}}$ are respectively the unique copulas equal to $A$ and $B$ according to Sklar. In practice, variables are on different supports (e.g., the amplitude of signals for $\delta$-frequency band versus the one for $\beta$-frequency band) and finding a finite grid of values $u$ and $v$ containing the support of both $\bf{X}$ and $\bf{Y}$ might be a tricky task. That is the reason why we standardize data into the $[0,1]$ interval (see how in Section \ref{changepoint}). 

Under a real equality in distribution for $C_{\bf{X}}$ and $C_{\bf{Y}}$, for $\tilde{u}\in [0,1]$ and $\tilde{v}\in [0,1]$ standardized versions of $u$ and $v$ being vectors of sufficiently large dimension, the statistic $D_{\bf{\tilde{X},\tilde{Y}}}$, where ${\bf{\tilde{X},\tilde{Y}}}$ are standardized versions of $\bf{X},\bf{Y}$, is nothing more than the bivariate version of the usual Kolmogorov-Smirnov statistic.

A remaining issue with Kolmogorov-Smirnov is that the validity of this statistic relies on the robustness of the distributions. However, due to the cardinality of the low-frequencies bands, estimating any parameter directly on these bands will lead to non-robust distributions. That is the reason why one has to use resampling techniques in order to obtain some distributions and then derive their parameters (e.g., mean and standard deviation).

\subsection{Block bootstrap for small frequency bands}

Due to the small cardinality of some frequency bands (i.e., those composed of a small quantity of single frequencies) such that $\Delta , \theta , \alpha$ or $\beta$ (e.g., the actual frequencies considered in the $\Delta$ band are $\{ 1,2,3,4 \}$ Hertz), any standard parametric inference methodology applied on the magnitude of the different Fourier frequencies within them, for a fixed epoch, will suffer from a lack of robustness. Indeed, with such small populations, any standard estimation (e.g., estimation of the parameters of the distribution) will lead to a statistic for which its variance with likely suffer from a lack of robustness. It is the reason why one has to use resampling methods while inferring distribution parameters in order to obtain a gain in robustness the variance of the estimators.

Let $X^{(r)}_\ell$ be the time-domain valued vector, of dimension $T$, for channel $\ell$ at epoch $r$. Computing straightforwardly the modulus of the Fourier transform, one obtains $\delta^{(r)}_{\Omega_\kappa, \ell}$, a vector whose cardinality might not be sufficiently large. We applied resampling techniques in order to obtain an empirial distribution of $\delta^{(r)}_{\Omega_\kappa, \ell}$. However, any naive use of bootstrap methods (\citet{efron1994}) will destroy the temporal structure among the $T$ observations of $X^{(r)}_\ell$. For this reason we apply the moving block bootstrap (see \citet{politis1994} or \citet{radovanov2014}) which preserves the temporal structure of the time series within an epoch. Here, we define $M$ to be the number of blocks, each with $T/M$ observations. Thus, one gets the bootstrapped variables $X_\ell^{b,(r)}$ for $b=1,...,B$ the number of iterations. One remarks that in this work, bootstrapped observations are only used to estimate the parameters of the distributions of $\delta_{\Omega_\kappa}^{(r)}$, they are not directly used on any measure of the strength of the dependence between variables represented through Kendall's tau or coherence measure.

\subsection{Estimation of the distributions}
Still for a reason of data size of $\delta_{\Omega_\kappa, \ell}^{(r)}$, we decided to avoid any empirical or non-parametric estimation of the distribution of $\delta_{\Omega_\kappa, \ell}^{(r)}$. As shown in \citet{brockwell2013}, the asymptotic distribution of the periodogram of a time series follows an exponential distribution with mean $\lambda$ equals to the spectrum. By some algebraic manipulations, we show in Appendix \ref{appendixRay} that the square root of an exponential distribution follows a Rayleigh distribution of parameter $1/\sqrt{2\lambda}$. Note that Rayleigh is a special case of the generalized Gamma distribution. Since the generalized gamma distribution is a model with three parameters (which allows room for computational bias in their estimation due to the idiosyncrasies of data for some frequency bands), we decided to use two-parameter models of that family to infer the distribution of $\delta_{\Omega_\kappa, \ell}^{(r)}$ in order to reduce inferential bias due to the small data size as well as to increase computational speed in the inferential process. Thus, we compared the likelihood of fitting a gamma distribution versus the one of fitting a two-parameters Weibull distribution to the LFP data, on all channel. Hence, with the help of an information criterion ($BIC$ - see Section \ref{copulaSel}), we decided to use the gamma distribution to model $\delta_{\Omega_\kappa, \ell}^{(r)}$. In the rest of this paper, we adopt the notation $\Gamma^{(r)}_{\Omega_\kappa, \ell}$ to denote the estimated distribution of the variable $\delta_{\Omega_\kappa, \ell}^{(r)}$ where the parameters $(\nu, \iota)$ are estimated by the maximum likelihood estimators $(\hat{\nu}, \hat{\iota})$. 

\subsection{Selection of a copula model} \label{copulaSel}

The copula-based algorithms to detect changes in brain signals, which are presented in this paper, can be fit using various types of copula functions. Among the most common ways of model selection from a wide set of possible types of copulas, we find those based on an information criterion. For instance, Akaike Information Criterion (AIC,  \cite{akaike1998}), Bayesian Information Criterion (BIC, \cite{schwarz1978}) or Copula Information Criterion (CIC, \citet{gronneberg2014}) are some of these possibilities. In this paper, to reduce the computational aspect of the algorithms and because only some slightly differences has been shown to exist between AIC and CIC (\citet{jordanger2014}), we use AIC to select all the copula models.

The range of copula models to consider for such a methodology is arbitrary. In this paper, in an attempt to avoid any numerical issues/misscomputations while computing the differences between some copula models (e.g., the difference between a normal copula and a Gumbel copula might be very high for border values due to their divergent behavior in these areas), we confined our choice only to the Archimedean family of copulas. We made this choice based on the flexibility of that family: elliptical copulas exhibit always a radial symmetry, which is not the case of the Archimedean copulas; furthermore Archimean copulas allow easily to model skewed distributions with non-symmetric tails. Thus, the panel of considered copulas was restricted to: independent, Clayton, Gumbel, Frank, Joe and rotated Joe ($180$ degrees) copulas (see \citet{cech2006} for more about rotated copulas).

\subsubsection{Effect of misspecification}

In this work, we suggest to limit the panel of available copula models to $6$ types of copulas from the Archimedean family. Let $\breve{C}^{(r,r')}_{\ell, \Omega_\kappa}(u,v; \breve{\theta})$, $r=1,...,R$ be the true copula (with its true parameter $\breve{\theta}$) which is maybe or not in our selection panel, and $C^{(r,r')}_{\ell, \Omega_\kappa}(u,v; \bar{\theta})$ be the one selected using AIC (or any other method based on likelihood information) with its parameter. Then, we express the Kullback-Leibler Information Criterion (KLIC) by
$$\text{KLIC}
=\int_0^1 \int_0^1 \text{log} \left( \frac{\breve{c}^{(r,r')}_{\ell, \Omega_\kappa}(u,v; \breve{\theta})}{c^{(r,r')}_{\ell, \Omega_\kappa}(u,v;\bar{\theta})} \right) d\breve{C}^{(r,r')}_{\ell, \Omega_\kappa}(u,v; \breve{\theta}). $$

If $\breve{C}^{(r,r')}_{\ell, \Omega_\kappa}(u,v; \breve{\theta})$ is in the panel of considered copulas choose based on a likelihood-based criterion, then indeed $\breve{C}^{(r,r')}_{\ell, \Omega_\kappa}(u,v; \breve{\theta})=C^{(r,r')}_{\ell, \Omega_\kappa}(u,v; \bar{\theta})$ and the KLIC will equal $0$. Otherwise, concerning the copula structure, as long as the real copula function $\breve{C}^{(r,r')}_{\ell, \Omega_\kappa}$ is unknown, it is not realistic to give a value of KLIC. However, we can minimize this criterion using a panel of flexible and rich possible families of copulas. Nevertheless, under a misspecified model, it is possible that the equivalent of $\breve{\theta}$ for the selected model does not exist. However, a pseudo-true parameter $\bar{\theta}$ exists. From \citet{white1982}, under conditions of continuity and measurability (which are met in this paper by LFP data) an estimator $\hat{\theta}_k$ of $\bar{\theta}$ obtained by a maximum likelihood estimator computed from the misspecified model will be, as $n=\text{card}(\Omega_\kappa) \to \infty$ (which is the case for Gamma band in the experimental setting described below, but might be the case for all frequency bands when there are enough time points within each epoch), consistent.

%Section presenting the first application:
\section{First application: Detecting a changepoint in across-epochs correlation over a frequency band, for a single channel}\label{changepoint}
\begin{figure}
\begin{center}
\includegraphics[scale=0.28 ]{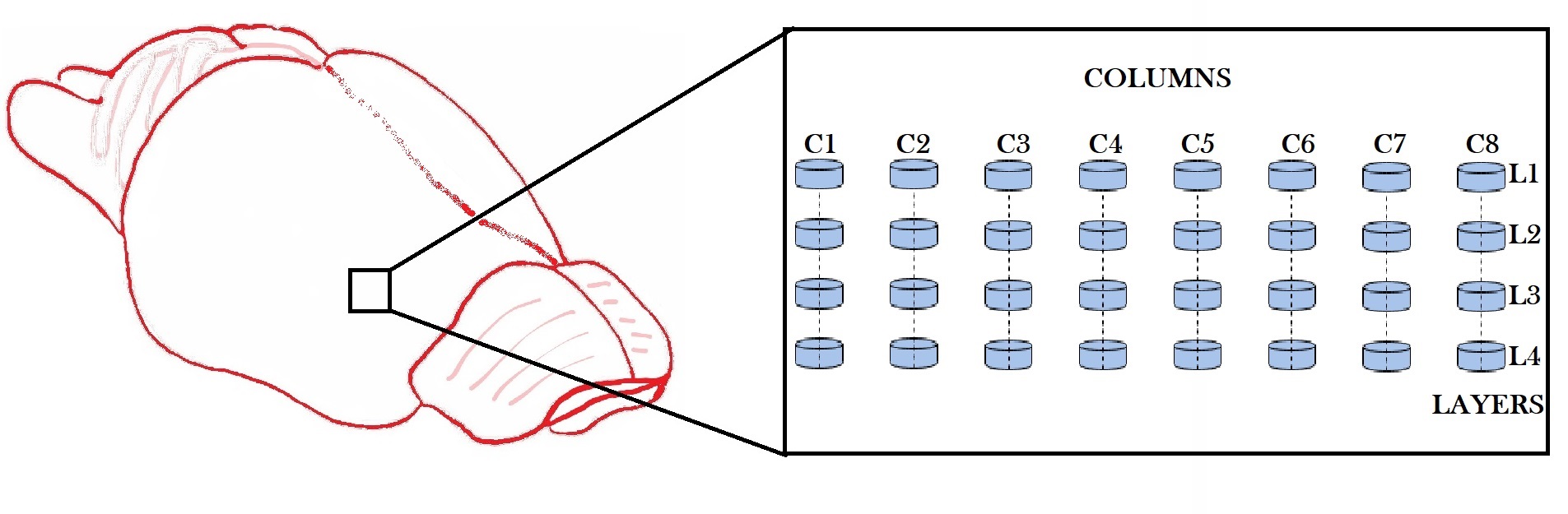}
\caption{Placement of the $32$ electrodes on the cortex of the rat. There are $4$ layers (having a different depth in the cortex: respectively $300\mu m$, $700\mu m$, $1100\mu m$ and $1500\mu m$) and each layer has $8$ electrodes. For details, see \citet{wann2017}. }
\label{RatBrain}
\end{center}
\end{figure}

To illustrate the pertinence of the dependence issues related to brain signals for certain frequency bands, here we use experimental data from \citet{wann2017} on local field potentials measured on rats' cortex. To summarize that experiment, local field potentials were recorded from $32$ microelectrodes placed on $4$ cortical layers (each with $8$ electrodes). This setup is illustrated in Figure \ref{RatBrain}. On these $32$ microelectrodes (channels), using insulated stainless steel wire electrodes, data have been recorded for 5 minutes where each second represents a single epoch which consists of $T=1000$ time points. After these five minutes, a stroke have been mechanically induced using an hemostat clamp on the brain artery located on the second column of electrodes (from the left) recording microelectrodes $2,10,18$ and $26$. Then, for five minutes, data, divided in the same way as the previous five minutes, has been recorded. 

\begin{figure}
\begin{center}
\includegraphics[scale=0.48]{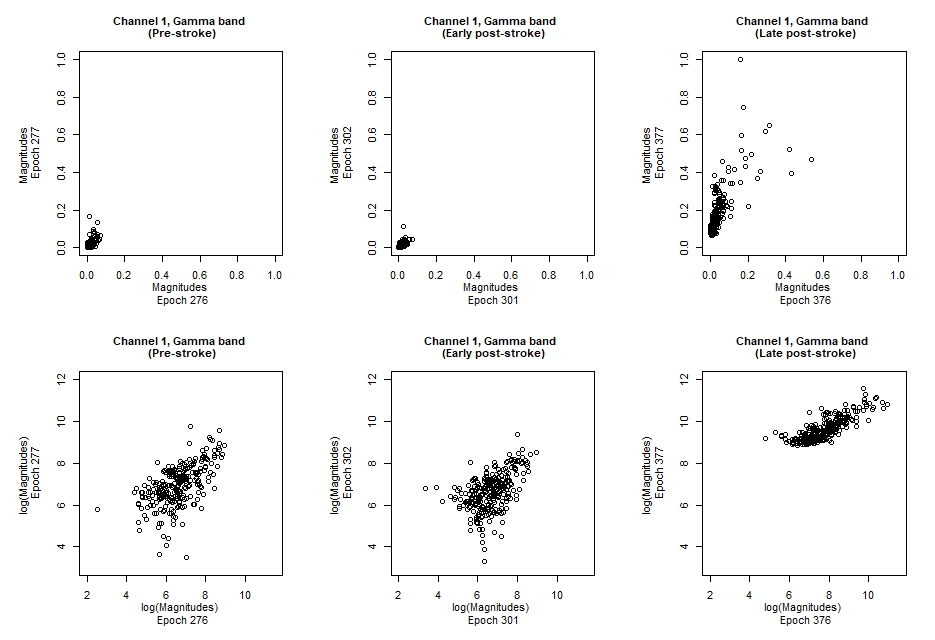}
\caption{Changes in between-epoch dependence for microelectrode (channel) $1$, for rat id $141020$. \textbf{First row: }Plot of the rescaled magnitudes into unit interval (as used in the changepoint detection algorithm) for pre-stroke \textsl{(left)}, early post-stroke \textsl{(middle)} and late post-stroke \textsl{(right)}. \textbf{Second row: }Plot of the log-scaled magnitudes (for visualization purpose) for pre-stroke \textsl{(left)}, early post-stroke \textsl{(middle)} and late post-stroke \textsl{(right)} .}
\label{scatterplotCh1}
\end{center}
\end{figure}

\begin{figure}
\begin{center}
\includegraphics[scale=0.35]{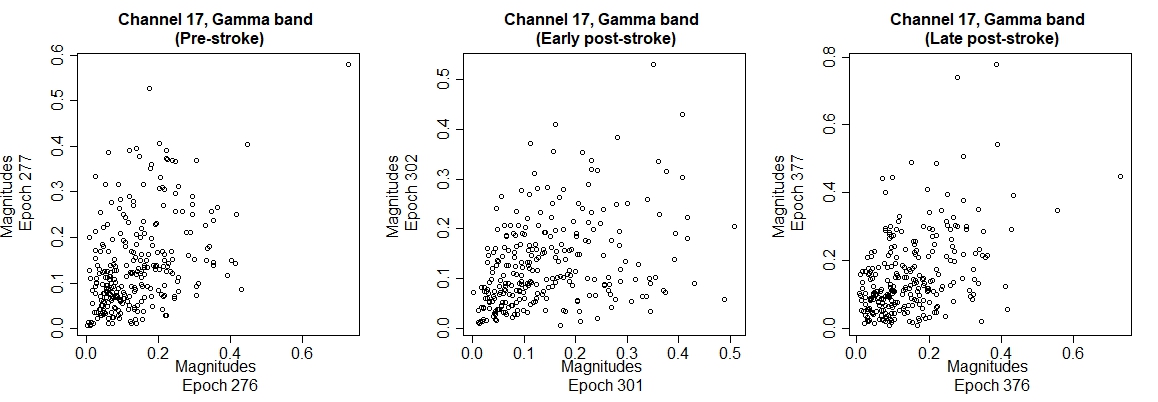}
\caption{Changes in between-epoch dependence for microelectrode $17$, for rat id $141020$. Plot of the rescaled magnitudes into unit interval for pre-stroke \textsl{(left)}, early post-stroke \textsl{(middle)} and late post-stroke \textsl{(right)} .}
\label{scatterplotCh17}
\end{center}
\end{figure}

Our interest in this section is to identify the epoch $r^{*}$ where the dependence between successive epochs $r^{*}-1$ and $r^{*}$ differ from the dependence between epochs $r^{*}$ and $r^{*}+1$. Hence, we are interested to identity that epoch $r^{*}$ for each microelectrode for all the $\Omega_\kappa \in \{ \Delta, \theta, \alpha, \beta, \gamma \}$ frequency bands. With LFP data, the dependence between $\delta_{\ell,\Omega_\kappa}^{(r)}$ and $\delta_{\ell,\Omega_\kappa}^{(r+1)}$ (no matters if these epochs are considered as a changepoint or not) exhibit frequently complex structure. For example, on Figure \ref{scatterplotCh17}, one sees (for rat id $141020$) for microelectrode (channel) $17$ some structures where the magnitudes of the Fourier coefficients for two successive epochs are highly dependent in their lower tail, and become more and more independent as they one moves toward their higher tail. This particular structure is easily representable through a copula function (Clayton copula will be considered in this case), but is not through any linear correlation (specifically coherence in the spectral domain) structure. For channel $1$ (Figure \ref{scatterplotCh1}) , one notices, still for the same rat, that the dependence structure change with stroke: for example, one notices the difference between the upper tails. This difference gets more and more obvious as we are looking for dependence between these magnitudes for epochs which get closer to the temporal interval: epochs $375$ to $380$. One notices that data in the first row are the one used in our algorithm. However, for a visualization purpose, data are log scaled in the second row in order to respond to skewness of large magnitude values. 

%\begin{figure}
%\begin{center}
%\includegraphics[scale=0.70]{Fig4_17.jpeg}
%\caption{Scatterplot of the dependence for the magnitudes of the Fourier coefficients in two different scales for two different channels. \textbf{First row: \textit{Pre-stroke dependence:} }\textsl{Left: } Dependence of log($\delta_{1,\gamma}^{(276)}$) vs log($\delta_{1,\gamma}^{(277)}$), \textsl{Right: } Dependence of the standardized (in the unit inverval) versions of $\delta_{17,\gamma}^{(276)}$ vs $\delta_{17,\gamma}^{(277)}$. \textbf{Second row: \textit{Early post-stroke dependence:} }\textsl{Left: } Dependence of log($\delta_{1,\gamma}^{(301)}$) vs log($\delta_{1,\gamma}^{(302)}$), \textsl{Right: }Dependence of the standardized versions of $\delta_{17,\gamma}^{(301)}$ vs $\delta_{17,\gamma}^{(302)}$. \textbf{Third row: \textit{Late post-stroke dependence}: }\textsl{Left: } Dependence of log($\delta_{1,\gamma}^{(376)}$) vs log($\delta_{1,\gamma}^{(377)}$), \textsl{Right: }Dependence of the standardized versions of $\delta_{17,\gamma}^{(376)}$ vs $\delta_{17,\gamma}^{(377)}$.}
%\label{scatterplotCh17}
%\end{center}
%\end{figure}

For this data, the expected major changepoint is $r^{*}=301$ which is the stroke onset. It is likely too that other changepoints would be observed after the stroke. However, for some biological issues, the peak of this observation might be delayed between the $375$$-th$ and the $380$$-th$ epoch  (from $75$ to $80$ seconds after the stroke) for a majority of the frequency channels, on most of their frequency bands. We observe that this $5-$seconds activity window is subject to change in function of the rat on which experiment is conducted. We note that the way we segregated epochs (changepoint vs stable epoch) is based on the empirical setting presented in \ref{empiricalSetting}.

\begin{figure}[!htbp]
\begin{center}
\includegraphics[scale=0.60 ]{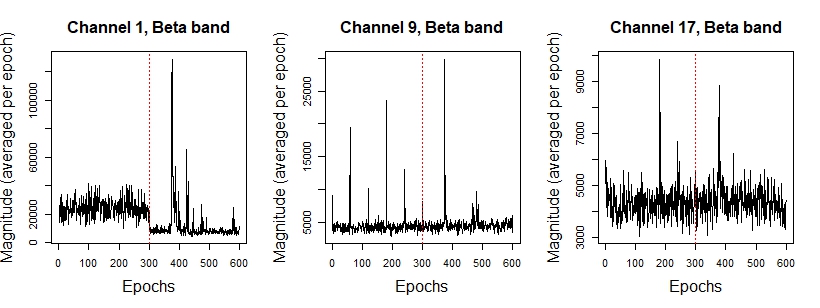}
\caption{Three different patterns in the regime of $\delta_{1, \beta}^{(1:600)},\delta_{9, \beta}^{(1:600)}$ and $\delta_{17, \beta}^{(1:600)}$. Red dotted lines represent the moment when the stroke is artificially induced.}
\label{threePat}
\end{center}
\end{figure}

We remark that, without regard to the frequency band, mainly three patterns are present in the regime of $\delta_{\ell, \Omega_\kappa}^{(1:600)}=[\delta_{\ell, \Omega_\kappa}^{(1)},...,\delta_{\ell, \Omega_\kappa}^{(600)}]$ with rat id $141020$. An interesting fact is that even if the location of the clamped artery is on column $2$, these three patterns are observed on column $1$. They are respectively microelectrodes (channels) $1$, $9$ and $17$. Figure \ref{threePat} exhibits the averaged amplitude (per epoch) for each one of these microelectrodes. The results for these three microelectrodes (for the five frequency bands) are presented as these are representative of our methodology.  

\vspace{0.5em}
\hrule\vspace{0.3em}
{ \noindent \textbf{\underline{ALGORITHM 1:} Detection of a changepoint over many epochs, for a particular microelectrode (channel) and a given frequency band}}\hrule \vspace{0.5em}
{\noindent \textbf{for} (epochs $r=1$ to $r=600$) \newline
1: \hspace{0.3in} Standardize (scale data into $[0,1]$ interval) such that \newline \hspace*{1in}
$\tilde{\delta}_{\ell, \Omega_\kappa}^{(r)}=(\delta_{{\ell, \Omega_\kappa}^{(r)}}-\min(\delta^{(1:600)}_{\ell, \Omega_\kappa}))/(\max(\delta^{(1:600)}_{\ell, \Omega_k})-\min(\delta^{(1:600)}_{\ell, \Omega_\kappa}))$. 
\newline
2: \hspace{0.3in} Apply the moving block bootstrap (to conserve the temporal structure inside \newline
     \hspace*{0.5in} data,  see Section \ref{section2}) by sampling on $X_\ell^{(r)}$ to obtain robust\newline
     \hspace*{0.5in} estimations of the shape $\nu$ and the rate $\upsilon$ values of a Gamma \newline
     \hspace*{0.5in} distribution and fit its cdf $\Gamma^{(r)}_{\ell,\Omega_\kappa}$ with parameters $(\nu^{(r)}_{\ell,\Omega_\kappa}, \upsilon^{(r)}_{\ell,\Omega_\kappa})$ \newline
\textbf{end } \newline
\textbf{for $r=1,...,599$} \newline
3: \hspace{0.3in} Compute Kendall's tau between $\delta_{\ell, \Omega_\kappa}^{(r)}$ and $\delta_{\ell, \Omega_\kappa}^{(r+1)}$. \newline
4: \hspace{0.3in} Among a predefined panel of parametric copulas, select using AIC the \newline
		 \hspace*{0.5in}most suitable copula model and using inverse Kendall's\newline
		 \hspace*{0.5in} tau method, estimate the corresponding copula dependence parameter. \newline
		 \hspace*{0.5in}This copula is noted $C^{(r,r+1)}_{\ell, \Omega_\kappa}\left( \Gamma^{(r)}_{\ell,\Omega_\kappa}(u),\Gamma^{(r+1)}_{\ell,\Omega_\kappa}(v) \right) \forall (u,v) \in [0,1]\times [0,1]$ \newline
\textbf{end } \newline
\textbf{for $r=2,...,599$} \newline
5: \hspace{0.3in} Compute all the bivariate Kolmogorov-Smirnov statistic $D(u,v)$ between \newline
		 \hspace*{0.5in} copulas $C^{(r-1,r)}_{\ell, \Omega_\kappa}\left( \Gamma^{(r-1)}_{\ell,\Omega_\kappa}(u),\Gamma^{(r)}_{\ell,\Omega_\kappa}(v) \right)$ and $C^{(r,r+1)}_{\ell, \Omega_\kappa}\left( \Gamma^{(r)}_{\ell,\Omega_\kappa}(u),\Gamma^{(r+1)}_{\ell,\Omega_\kappa}(v) \right)$ \newline
\textbf{end }\newline
\textbf{Output:} Kolmogorov-Smirnov statistics above a threshold determined for a desired significance level are said to be related to changepoint epochs.
\vspace{0.5em} \hrule}

Algorithm 1 presents the proposed methodology to detect one or many changepoint(s) in the regime of a channel for a given frequency band. Some remarks are as follows. Firstly, in this algorithm, we adopt the tilde notation (e.g., $\tilde{\delta}_{\ell, \Omega_\kappa}^{(r)}$) to differentiate standardized variables from regular variables, but in this paper we assume all variables to be standardized after this step. Therefore,  we will no longer use this notation in order to simplify the text. Also, still in the same step, note that we use unique minimum and maximum over all the $600$ epochs range. It is crucial to use these standardization values in order to get comparable Kolmogorov-Smirnov statistics. Secondly, our panel of copulas consists only of Archimedean models in order to avoid potential problems of comparing widely different models. Finally, the thresholds for the Kolmogorov-Smornov statistics are determined empirically and the procedure is described in Section \ref{empiricalSetting}.

\subsection{Empirical thresholds for Kolmogorov-Smirnov statistics}\label{empiricalSetting}
The goal of this subsection is to determine empirical threshold(s) for the bivariate Kolmogorov-Smirnov statistics, which will be used in order to test for a change in the auto-correlation between succeeding epochs. The determination of a theoretical threshold under the conditions on data used in this paper is a work in progress. Thus, as it is an explanatory work where we want to illustrate the potential of our methodology, we establish from some simulations these thresholds through two main scenarios of data generating processes (DGPs). 

The overall idea in all these DGPs is to simulate two or more time series (with moderate noises) in a given DGP, from a latent signal derived from an autoregressive process. The reason justifying to simulate many different series in each simulation is to explore the effect of various latent signals with our copula-based algorithm.

We mention that the hypothesis we are considering to establish a significant threshold are the equivalence of $C^{(r-1,r)}_{\ell, \Omega_\kappa} \equiv C^{(r,r+1)}_{\ell, \Omega_\kappa}$, $r=2,...,R-1$ under the null hypothesis against the hypothesis of non-equivalence under the alternative one. It can be rewritten as:
$$
\begin{cases}
      H_0: \left| C^{(r-1,r)}_{\ell, \Omega_\kappa}(u,v) - C^{(r,r+1)}_{\ell, \Omega_\kappa}(u,v) \right| = 0
     & \forall (u,v) \in [0,1] \times [0,1];
		\\[4pt]
       H_1: \left| C^{(r-1,r)}_{\ell, \Omega_\kappa}(u,v) - C^{(r,r+1)}_{\ell, \Omega_\kappa}(u,v) \right| > 0 
      & \text{for some } (u,v) \in [0,1] \times [0,1].
\end{cases}
$$

Thus, setting up an experimental-based threshold that provides a critical value to test these hypotheses at a significance level $\bar{\alpha}$ (in order to avoid confusion with $\alpha$, a frequency band) is our challenge here. We remark that we fixed our risk of type 1 error to $\bar{\alpha}=1\%$. 

\subsubsection{Deriving the empirical thresholds under the null hypothesis}
\paragraph{DGP 1}
In this DGP, we simulated two different scenarios where, for each scenario, we simulated $R=100$ epochs with $T=1000$ timepoints per epoch. The first scenario follows a stationary $AR(1)$, then the second one follows a similar $AR(1)$ process where we added a constant. The simulations setting is, for $t=1,...,T$:
\begin{itemize}
\item $Z_{t,A}^{(r)}= 0.9X_t^{(r)}+\epsilon_t^{(r)}$ where $X_t^{(r)} \sim \text{AR}(1) \text{ of parameter }\phi=0.9$, $\epsilon_t^{(r)} \sim \mathcal{N}(0,0.1)$ for $r=1,...,R$
\item $Z_{t,B}^{(r)}= 1+0.9X_t^{(r)}+\epsilon_t^{(r)}$ where $X_t^{(r)} \sim \text{AR}(1) \text{ of parameter }\phi=0.9$, $\epsilon_t^{(r)} \sim \mathcal{N}(0,0.1)$ for $r=1,...,R$
\end{itemize}
 
We computed in both cases the bivariate Kolmogorov-Smirnov statistics, $D(u,v)$, between each consecutive pairs of copulas $C^{(r-1,r)}$ and $C^{(r,r+1)}$, $r=2,...,199$. These statistics are plotted for each of the three frequency bands on Figure \ref{KSdgp1}.  We remark that this DGP is considered being a basic simulation model. The goal here is to establish an empirical distribution of the Kolmogorov-Smirnov statistic under the null hypotheses and to identify the $99-th$ percentile which will serve as the threshold that satisfies $\mathbb{P}$(Type I error)$ = \bar{\alpha} = 0.01$.

\begin{table}[h]
\begin{center}
\begin{tabular}{|l||*{1}{c|}}

    \hline
     \textbf{Frequency band} & Threshold  \\
		                         & for $\bar{\alpha}=1\%$  \\
    \hline \hline
		Delta band $(\Delta)$ &  $D(u,v) >  0.0102$  \\ \hline
		Theta band $(\theta)$  &    $D(u,v) > 0.0451$ \\ \hline
    Alpha band $(\alpha)$ &  $D(u,v) > 0.0090$ \\ \hline
		Beta band $(\beta)$  &   $D(u,v) > 0.0038$    \\ \hline
		Gamma band $(\gamma)$  &  $D(u,v) > 0.0048$  \\ \hline
		  \end{tabular}
	\caption{DGP 1: threshold on the Kolmogorov-Smirnov statistics for a significance levels of $\bar{\alpha}=1\%$.   }
	\label{tableDGP1}
\end{center}
\end{table}

\paragraph{DGP 2}
The second DGP is based on some AR(2) processes. There main idea for this DGP is to analyze time series generated from multiple latent signals, where the time series used is tributary of the frequency band on which is performed the analysis. We notice the stationarity here across epochs (i.e., dependence between successive epochs does not change).

The principle is that latent signals from six AR(2) processes are observed for $100$ epochs. Thus, the six latent signals are: $X_{t,i} \sim AR(2), i=1,...,6$ with polynomial function whose roots are complex-valued with  respectively, for each latent signal, phases $p_1= \pm 4/T \cdot 2\pi$, $p_2=\pm 6/T \cdot 2\pi$, $p_3= \pm 9/T \cdot 2\pi$, $p_4= \pm 13/T \cdot 2\pi$, $p_5=\pm 15/T \cdot 2\pi$, and $p_6= \pm 150/T \cdot 2\pi$; for $t=1,...,1000$. Thus, the spectra of these latent signals are concentrated on the phases of each one of the bands of interest. Our simulation setup, for $t=1,...,T$ and for $r=1,...,100$, is:
\begin{itemize}
\item $Z_{t,i}^{(r)}=X_{t,i}^{(r)}+\epsilon_t^{(r)},$ with noise $\epsilon_t^{(r)} \sim \mathcal{N}(0, 0.1 \sigma_{X_{t,i}^{(r)}}); \text{ for }i=1,...,6.$
\end{itemize}
 
Table \ref{tableDGP2a} presents the thresholds obtained for a significance value of $\bar{\alpha}=1\%$. Hence, for Delta band for example, based on these simulations, assuming the null hypothesis true, the epochs related any Kolmogorov Smirnov statistic valued greater than $0.0149$ will be considered as a changepoint in the dependence structure.

\begin{table}[h]
\begin{center}
\begin{tabular}{|l||*{1}{c|}}

    \hline
     \textbf{Frequency band}   & Threshold for $\bar{\alpha}=1\%$ \\
    \hline \hline
		Delta band $(\Delta)$  & $D(u,v) > 0.0149$ \\ \hline
		Theta band $(\theta)$  &  $D(u,v) > 0.0625$ \\ \hline
    Alpha band $(\alpha)$  & $D(u,v) > 0.0101$ \\ \hline
		Beta band $(\beta)$  & $D(u,v) > 0.0050$ \\ \hline
		Gamma band $(\gamma)$  & $D(u,v) > 0.0103$ \\ \hline
		  \end{tabular}
	\caption{DGP 2: Threshold on the Kolmogorov-Smirnov statistics for a significance value of $\bar{\alpha}=1\%$.}
	\label{tableDGP2a}
\end{center}
\end{table}

\subsubsection{Empirical threshold}
We conclude that each frequency band has its own threshold. We note that these critical values are not based on a theoretical development but they are empirically based on an explanatory work. Thus, they are tributary to the way we infer the copulas in our code as well as the way that we compute Kolmogorov-Smirnov statistics over a bidimensional grid of evaluation points. However, as our methodology and our code remain the same to analyze LFP data, these threshold are a reliable way to determine changepoint(s) in the rats brain activity. Hence, as all the thresholds determined in DGP 2 are more conservative than the one in DGP 1, we will consider the latter (see table \ref{tableDGP2a}) in our local field potential of a rat study. \

\subsubsection{Illustration of the power of the test, under $H_1$}

To assess the power of the test (under the alternative hypothesis), we decided to retake both DGPs from the last section and to combine the simulations settings in the same scenario. For example, for DGP $1$, the simulation setting now becomes a scenario of $200$ epochs such that we observe $Z_{t,A}^{(r)}$ for $r=1,...,100$ and then $Z_{t,B}^{(r)}$ for $r=101,...,200$. Thus, we observe two consecutive stationary series where the dependence between epochs does not change in the first half ($r=1,...,100$), the one between epochs does not change too in the second half ($r=101,...,200$), and a changepoint is expected between both stationary blocks.

For each DGP, we collect all the Kolmogorov-Smirnov values and verify that the known changepoint(s) (location between two consecutive series) are above the threshold. For example, as the series $Z_{t,A}^{(r)}$ for epochs $r=1,...100$ and the series $Z_{t,B}^{(r)}$ for epochs $r=101,...,200$ are different, we will expect to detect a changepoint; which means that $|C^{(99,100)}-C^{(100,101)}|\geq \bar{\alpha}$ as well as $|C^{(100,101)}-C^{(101,102)}|\geq \bar{\alpha}$ for a significance level $\bar{\alpha}$ determined empirically; which leads to $2$ values above the thresholds.

We present, in appendix \ref{appendixThresh}, for three frequency bands, the Kolmogorov-Smirnov statistics for DGP $1$ when, for a scenario of $200$ epochs, for $t=1,...,1000$, the setup becomes $Z_{t,A}^{(r)}$ for $r=1,...,100$ and $Z_{t,B}^{(r)}$ for $r=101,...,200$. We remark on Figure \ref{KSdgp1} that the Kolmogorov-Smirnov statistics related to epochs $99$ and $101$ are above the threshold line.

\begin{figure}
\begin{center}
\includegraphics[scale=0.55 ]{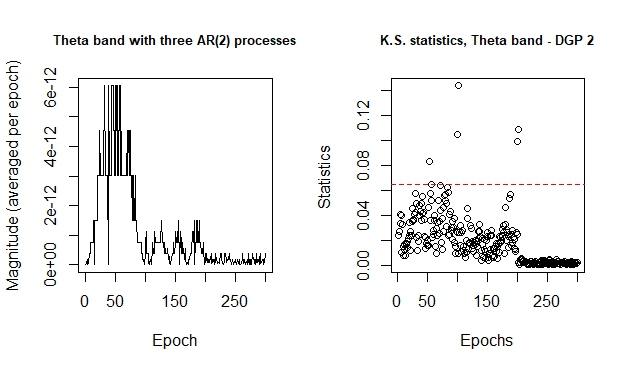}
\caption{DGP 2: Illustration of the power of the test, under the alternative hypothesis, for three observed series and for theta band. \textbf{Left:} Plot of the magnitude of Fourier coefficients (averaged per epoch) for the three consecutive series. \textbf{Right:} Kolmogorov-Smirnov statistics obtained and the empirically derived threshold for $\bar{\alpha}=1\%.$}
\label{DGP2powerTheta}
\end{center}
\end{figure}

Concerning DGP $2$, we decided to combine three of the six series: $Z_{t,2}^{(r)} \text{ for }r=1,...,100, Z_{t,5}^{(r)} \text{ for }r=101,...,200 \text{ and }Z_{t,6}^{(r)} \text{ for }r=201,...,300$. We show the results in Figure \ref{DGP2powerTheta} for theta band.

\subsection{Changepoints observed on LFP}

\begin{figure}
\begin{center}
\includegraphics[scale=0.65 ]{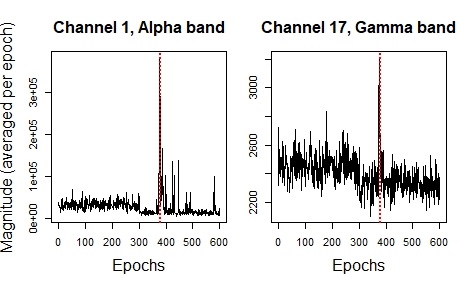}
\caption{Examples (for two microelectrodes, for rat id $141020$) of detection of the epochs related to a changepoint in the dependence structure (for $\bar{alpha}-1\%$) using the copula-based method, represented by the vertical red dashed lines. \textbf{Left:} For channel 1, alpha band. \textbf{Right:} For channel 17, gamma band.}
\label{changepoints}
\end{center}
\end{figure}

To verify the validity of our method which has been applied to LFP data (results for rat id $141020$ are presented here), we decide to use also an estimator to detect changepoint by pruned objectives (see \citet{james2015}), and to verify if changepoint(s) detected by both methods concord. Figure \ref{changepoints} presents for two channels the detected changepoints, which are stated on table \ref{changeEpochs}, for a significance level of $\bar{\alpha}=1\%$. We remark that most of the time, epochs $374$ to $379$ are significant changepoints.

\begin{table}{\footnotesize
\begin{tabular}{|l||*{1}{c|c|}} 

    \hline
     \textbf{Channel and} & Copula-based &  Condordance with    \\
		 \textbf{Frequency band}    & algorithm & James algorithm \\
    \hline \hline
    Channel 1, $\Delta$ &$374,375,376,377,378,399,425$ &  Yes\\ 
		Channel 1, $\theta$ &$\*$(\textit{25 changepoints have been detected}) & Yes \\ 
		Channel 1, $\alpha$ &$375,376,377,378$ &  Yes \\
		Channel 1, $\beta$ &$374,375,376,378,379,380,387$ &  Yes \\
		Channel 1, $\gamma$ &$374,375,376,378$ & Yes \\
		\hline
		Channel 9, $\Delta$ &$59,60,61,62,179,180,181,182,239,240$ & Yes \\ 
		Channel 9, $\theta$ &$121,372,373$ & Yes  \\
		Channel 9, $\alpha$ & $214, 309, 372, 373, 374, 375, 376, 480$ & No \\
		Channel 9, $\beta$ &$56,372,373,374,375,376$  & Yes\\
		Channel 9, $\gamma$ &  $60, 180, 182, 372, 373, 374, 375, 376$ & No  \\
		\hline
	  Channel 17, $\Delta$ &$179, 180, 181, 344, 345, 374, 375, 376, 377, 378, 399, 424, 425, 426, 524, 525$ & No \\
		Channel 17, $\theta$ & $18, 88, 191, 215, 339, 373, 378, 390, 446, 467, 473, 563$ & Yes \\
		Channel 17, $\alpha$ &$\*$(\textit{30 changepoints have been detected}) & Yes \\
		Channel 17, $\beta$ & $\*$(\textit{91 changepoints have been detected}) & Yes \\
		Channel 17, $\gamma$ &$375, 376, 377, 378$ & No \\
		\hline 
		   \end{tabular}}

	\caption{Epochs considered as changepoints using the copula-based algorithm (with a threshold of $\bar{\alpha}=1\%$), for rat id $141020$.  }
	\label{changeEpochs}

\end{table}

%Section presenting the second application: comparing dep of the whole 1:300 epochs, to the one of the whole 301:600
\section{Second application: Comparing dependence prior to and post induced stroke}\label{prepost}

Our goal here is to compare the spectral dependence of the magnitude of Fourier coefficients pre-stroke versus the one post-stroke (i.e., to compare for a fixed microelectrode - understand ''a fixed channel''- and a fixed frequency band if there is a change in the entire structure of dependence among the $300$ epochs before the stroke versus after).

To do so, consider a given frequency band $\Omega_\kappa$, $\kappa=1,...,Q$. We define the multivariate matrices $ \delta_{\ell,\Omega_\kappa}^{(1:300)}=[\delta_{\ell,\Omega_\kappa}^{(1)},...,\delta_{\ell,\Omega_\kappa}^{(300)} ]$ and $ \delta_{\ell,\Omega_\kappa}^{(301:600)}=[\delta_{\ell,\Omega_\kappa}^{(301)},...,\delta_{\ell,\Omega_\kappa}^{(600)} ]$ (two matrices of dimension $\text{card}(\Omega_\kappa) \times 300$) as two single structures of the dependence. Using straightforwardly a single parametric copula in each case would be an enormous mistake. In fact, the parameter(s) of any Archimedean copula is too general to represent at the same time both the dependence measure between $\delta_{\ell,\Omega_\kappa}^{(1)}$ and $\delta_{\ell,\Omega_\kappa}^{(2)}$ and the dependence measure between $\delta_{\ell,\Omega_\kappa}^{(r)}$ and $\delta_{\ell,\Omega_\kappa}^{(r+1)}; r=2,...,299$. That's the reason why we propose here to use \textit{vine copulas} (for information, see see \citet{bedford2002} and \citet{aas2009}) to represent the dependence between these sets of variables.

The principle of vine copulas is the representation of a multivariate copula as a nested network of bivariate copulas where each single copula is named a \textit{node} and each link between two nodes (defining the order of the copulas and their relations among themselves) is named an \textit{edge}. Each level of dependence in this nested network is named a \textit{tree}. For all type of vines, the first tree is always the set of copulas between the univariate nodes (variables) and for the following trees, the nodes are always conditionals to at least one variable.

In this paper, due to the temporal relation between the consecutive $\delta_{\ell, \Omega_\kappa}^{(r)}, r=1,...,600$, we assume the structures of dependence for the multivariate sets $ \delta_{\ell,\Omega_\kappa}^{(1:300)}$ and $ \delta_{\ell,\Omega_\kappa}^{(301:600)}$ being represented by drawable vine (D-Vine) copulas. In other words, we assume that the edge between any node in the first tree only link the consecutive variables $\delta_{\ell,\Omega_\kappa}^{(r)},\delta_{\ell,\Omega_\kappa}^{(r+1)}; r=1,...,599$.

Obviously, we cannot use a 300-variate version of the Kolmogorov-Smirnov statistic to compare $ \delta_{\ell,\Omega_\kappa}^{(1:300)}$ to $ \delta_{\ell,\Omega_\kappa}^{(301:600)}$. It is still computationally unfeasible. That's the reason why, in our work, we propose to adapt a test comparing two vine copulas models (see \citet{clarke2007})to our context consisting in determining any difference in the structure of dependence between them. We remark that this test is mainly used in the literature in a goodness of fit perspective of a vine structure given a set of data. Since $ \delta_{\ell,\Omega_\kappa}^{(1:300)}$ and $ \delta_{\ell,\Omega_\kappa}^{(301:600)}$ are two different set of data having the same dimensionality which does not need to be independent, it is appropriate to use it. 

The principle of that test is as follow. Let the ratios of the log-likelihhood for each Fourier frequency in the band under consideration $m_{i;\ell,\Omega_\kappa}=\text{log }\left\{ \frac{c^{(1:300)}_{\ell,\Omega_\kappa}({\bf{u}}_i|\theta^{(1:300)}_{\ell,\Omega_\kappa})}{c^{(301:600)}_{\ell,\Omega_\kappa}({\bf{u}}_i|\theta^{(301:600)}_{\ell,\Omega_\kappa})}  \right\} $,  $i=1,...,\text{card}(\Omega_\kappa)$ where $c$ stands for the density of a D-Vine copula function, $\theta^{(1:300)}_{\ell,\Omega_\kappa}, \theta^{(301:600)}_{\ell,\Omega_\kappa}$ the vectors of the copula parameters for each vine structure and ${\bf{u}}_i$ the vector of observations. Thus, if there is no difference between the vine copulas of the sets of variables $ \delta_{\ell,\Omega_\kappa}^{(1:300)}$ and $ \delta_{\ell,\Omega_\kappa}^{(301:600)}$, the ratios of the log-likelihhood $m_{i;\ell,\Omega_\kappa}$ should be uniformly distributed around zero and $50\%$ of them should be greater than 0 (for details and proof, see \citet{vuong1989}). Thus, we are testing for all $i=1,...,\text{card}(\Omega_\kappa)$: $$ \begin{cases} H_0:\mathbb{P}\left(  m_{i;\ell,\Omega_\kappa} > 0 \right)=0.5, \\ H_1:\mathbb{P}\left(  m_{i;\ell,\Omega_\kappa} > 0 \right)\neq 0.5. \end{cases}$$ Therefore, the statistic of test is $$\xi_{\ell,\Omega_\kappa}=\sum_{i=1}^{\text{card}(\Omega_\kappa)}\mathds{1}_{(0,\infty)}\left( m_{i;\ell,\Omega_\kappa} \right)$$ where $\mathds{1}$ stands for the indicator function. Then, under the null hypothesis, $\xi_{\ell,\Omega_\kappa} \sim \mathcal{B}\text{\textit{in}}(\text{card}(\Omega_\kappa),0.5)$ and we can interpreted the statistic of test such that the vine copula pre-stroke is statistically equivalent to the one post stroke if $\xi$ is not statistically different from $\mathbb{P}(m_{i;\ell,\Omega_\kappa} \leq 0) \times \text{card}(\Omega_\kappa) = 0.5\text{card}(\Omega_\kappa)$ for a given significance level. This test is known as Clarke's test \cite{clarke2007} and has been considered in most of the literature comparing two vine structures (see \citet{joe2011}).

\begin{figure}
\begin{center}
\includegraphics[scale=0.28]{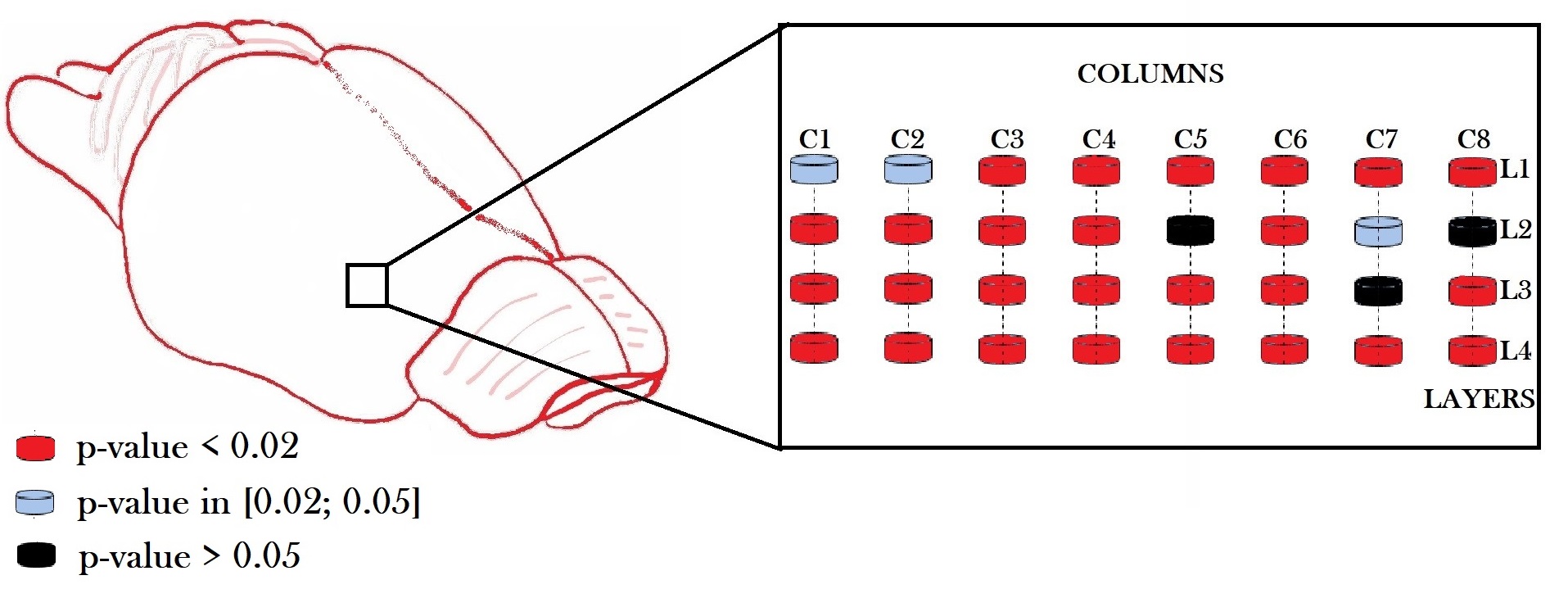}
\caption{Schema showing, for rat id $141020$, the p-values found on the $32$ channels for Gamma band.}
\label{ratbrainPrePost}
\end{center}
\end{figure}

We performed our version of that test on the gamma band for the whole LFP data set, as it is the one where we can visually observe on some channels aspects of change and on some other channels aspect of stability. We note that, in order to reduce noise, we truncated gamma band to $300$ $Hz$ such that $\gamma \in \{ 30, 300 \} Hz$. Figure \ref{ratbrainPrePost} shows, for rat id $141020$, in the way the electrodes are placed in the rat's brain the p-value obtained for each channel. In Appendix \ref{appendixPrePost}, one observes the statistics of test we obtained for each channel (second row) as well as these p-values for each one of the four rats. Microelectrodes in black suggest to accept $H_0$ for a significance level of $2.5\%$. Thus, under that significance level, we can say that for $\gamma$-band, there are no strong evidences of a change in the brain activity of the rat after the induced stroke for channels $1,2,13,14, 15,16 \text{ and }23$.

\section{Third application: Comparing the dependence behavior of two different channels for a given frequency band}\label{twoChan}

This section is in fact a brief note to show that one can apply the methodology from Section \ref{prepost} to compare, based on the dependence structure, if two different microelectrodes (brain channels), for a given frequency band, act similarly during all the regime of the experiment (i.e., during the $600$ epochs). Thus, we test exactly the same hypothesis but this time, $m_{i;\ell,\ell';\Omega_\kappa}$ is defined differently. For epochs $1$ to $600$, this log-likelihood ratio is defined by:
$$ 
m_{i;\ell,\ell';\Omega_\kappa}=\text{log }\left\{ \frac{c^{(1:600)}_{\ell,\Omega_\kappa}({\bf{u}}_i|\theta^{(1:600)}_{\ell,\Omega_\kappa})}{c^{(1:600)}_{\ell',\Omega_\kappa}({\bf{u}}_i|\theta^{(1:600)}_{\ell',\Omega_\kappa})}  \right\} 
$$
where $ i=1,...,\text{card}(\Omega_\kappa)$, and $\ell$ and $\ell'$ are obviously two different channels. We applied this test to the channels defined on the two first columns of microelectrodes in the rat brain (i.e., channels linked to microelectrodes $1,2,9,10,17,18,25$ and $26$; which means a total of $28$ possible combinations). We show the results for the gamma band, for the four experimental rats, in Appendix \ref{appendixPrePost} (Table \ref{changeChannels}). We remind that to not reject $H_0$, the Clarke's statistic of test must not be significantly different to $\text{card}(\Omega_\kappa)/2$. As the results are for gamma band, they should not be significantly different to $(300\text{Hz}-31{Hz})/2=135$ under $H_0$. That said, even if the Clarke's statistics are valued on a wide range from $0$ to $270$, one observes that these $8$ channels are considered being completely different on their whole regime for a significance level: the p-values are always lower than $0.0001$ for all the $28$ possible combinations.

%Conclusion
\section{Conclusion}
This paper related some approaches to assess both the dependence and the information we can learn from that dependence (i.e. changepoint, change in a regime, etc.). By considering more complex structures than simply linear relations (e.g., linear correlation, coherence) to assess dependence between brain signals, we modeled the relations between these signals. We also proposed algorithms with which we determined if one can presume a change or not on these complex structures of dependence. Such a methodology aims to show his utility in the future because research about specific types of strokes like CVA gets more and more funded in order to do prevention in the society.

In closing, we address two potential criticisms of the proposed work. Firstly, we used only parametric copula models when it is true that in general, non-parametric models are more flexible to data. But in our context, the dimension of some frequency bands is not large enough to ensure the robustness of a nonparametric model as a parametric model might be. Secondly, the analysis was conducted only on four rats. It is true that data from many more rats will increase the power of the neurological conclusions. However, the work done here was an explanatory study of a copula-based approach for such data, and having to analyze data from many more rats will complexity the computational work. Many future research avenues are possible from what we did. One of these avenues is to study the impact of taking copulas on more than $2$ epochs while processing the iterative algorithm when studying a changepoint for a single brain channel. Indeed, it will allow to detect changes that occur for small windows of time instead that punctually. Another one is to study our copula-based approach the possible lagged dependence(s) between two different frequency bands.

\section*{Acknowledgements}
Ron D. Frostig was supported by the Leducq Foundation (grand 15CVD02).
\bibliographystyle{chicago}
\begin{flushleft}

\end{flushleft}
{\setstretch{1.5}
\bibliography{refJapS}
}

\begin{appendices}
\newpage
\section{Table of notation}\label{appendixNotation}
\begin{table}[!htbp]
\begin{center}
{\small
\begin{tabular}{l||*{1}{l}}
	\underline{\textbf{Notation}}                      & \underline{\textbf{Signification}} \\
     $\ell, \ell'$                      & index of a channel, $\ell, \ell'=1,...,d$ \\
		 $(r), (r')$                       & index of an epoch \\
		 $T$                         & number of time points for each epoch, assumed to be even \\
		                             &  \\
		$\bf{X}^{(r)}$               & $\bf{X}^{(r)}= [X_1^{(r)},...,X_d^{(r)}]' $matrix of size $T \times d$ containing the entire observations \\ & \hspace{2em} for epoch $r$ \\
		$X_\ell^{(r)}$               & vector, in time domain, of $T$ time points for channel $\ell$ at epoch $r$ \\
		                             &  \\
	  $f_{\ell,\omega_k}^{(r)}$    & Fourier transform of $X_\ell^{(r)}$ \\
		$\omega_k$                   & Fourier fundamental frequencies$:= k/T$ \\
		$\Omega_\kappa,\Omega_{\kappa'} $              & frequency band, $\kappa, \kappa'=1,...,Q$; (e.g. with $\Omega=\{\Delta, \theta, \alpha, \beta, \gamma$\}, $Q=5$) \\
		$\delta_{\omega_k}^{(r)}$   & $\delta_{\omega_k}^{(r)}= \left[ |f_{1,\omega_k}^{(r)}|,...,|f_{d,\omega_k}^{(r)}|\right]'$ \\
		$\delta_{\Omega_\kappa}^{(r)}$ & matrix of dim $\text{card}(\Omega_\kappa)\times d$ containing all the fundamental frequencies for a given\\  & \hspace{2em} band at a given epoch \\
		$ \delta_{\ell,\Omega_\kappa}^{(r:s)}$ & matrix of the variables $ [\delta_{\ell,\Omega_\kappa}^{(r)},...,\delta_{\ell,\Omega_\kappa}^{(s)} ], r,s \in \{1,...,R\}, r \leq s$ \\
		                              &    \\
	 $H_{(\ell,\Omega_\kappa)(\ell ', \Omega_\kappa ')}^{(\delta_{\ell,\Omega_\kappa}^{(r)},\delta_{\ell ',\Omega_\kappa '}^{(r)})}$ & joint cdf of $\delta_{\ell,\Omega_\kappa}^{(r)},\delta_{\ell ',\Omega_\kappa '}^{(r)}$, $\ell, \ell' = 1,...,d$, $\kappa, \kappa' =1,...,Q$, $r=1,...,R$ \\
	$H_{\ell,\Omega_\kappa}^{(r)}(\delta_{\ell,\Omega_\kappa}^{(r)})$  & marginal cdf for channel $\ell=1,...,d$, frequency band $\Omega_\kappa, \kappa=1,...,Q$ at \\ & \hspace{2em} epoch $r=1,...,R$   \\
	$C_{(\ell,\Omega_\kappa)(\ell ', \Omega_\kappa ')}^{(r,r')}$ & copula function between $\delta_{\ell,\Omega_\kappa}^{(r)}$ and $\delta_{\ell',\Omega_\kappa'}^{(r')}$
   \\
	$C_{\ell,\Omega_\kappa}^{(r,r')}$ & copula function between $\delta_{\ell,\Omega_\kappa}^{(r)}$ and $\delta_{\ell,\Omega_\kappa}^{(r')}$ \textit{(in case $\ell=\ell', \kappa=\kappa'$)} \\
	$c_{\ell,\Omega_\kappa}^{(r,r')}$ & density of the copula $C_{\ell,\Omega_\kappa}^{(r,r')}$ \\
	 & \\
	$\breve{C}^{(r,r')}_{\ell, \Omega_\kappa}$ & true copula function between $\delta_{\ell,\Omega_\kappa}^{(r)}$ and $\delta_{\ell,\Omega_\kappa}^{(r')}$ \\
	$\breve{\theta}$ & true copula parameter \\
	$\bar{\theta}$ & pseudo-truc copula parameter \\
	$\hat{\theta}_{{\bf K}}$ & maximum likelihood estimator of the true copula parameter \\
	$\hat{\theta}_\tau$ & estimator of the true copula parameter based on the inversion of the Kendall's tau \\
	$\Gamma^{(r)}_{\ell,\Omega_\kappa}$ & Gamma distribution fitted to $\Gamma^{(r)}_{\ell,\Omega_\kappa}$ \\
	$\nu, \upsilon$ & parameters of a Gamma distribution \\
	$\hat{\nu},\hat{\upsilon}$ & maximum likelihood estimators of $\nu,\upsilon$ \\
	$\theta^{(r:s)}_{\ell,\Omega_\kappa}$ & vector of the copula parameters of a vine structure  \\ & \hspace{2em} joining all the variables from $\delta_{\ell,\Omega_\kappa}^{(r)}$ to $\delta_{\ell,\Omega_\kappa}^{(s)}$, $r<s$ \\
	
                 & \\
		$\tilde{u},\tilde{v}$ & standardized version of the vectors $u,v$ in $[0,1]$ \\
		$\bf{\tilde{X}},\bf{\tilde{Y}}$ & standardized versions on $\bf{X},\bf{Y}$ \\
	$X_\ell^{b,(r)}$ & bootstraped version of $X_\ell^{(r)}$ at the $b$-th iteration\\
	$b$             & $b=1,...,B$ index of the iteration in the bootstrap process \\
	$M$             & number of blocks (bootstrap procedure) of size  $T/M$ \\
	    &   \\
			$D(u,v)$ & Kolmogorov-Smirnov statistic \\
	$m_{i;\ell,\Omega_\kappa}$ & ratio of the pointwise log-likelihoods pre-stroke over post-stroke \\
	$\xi_{\ell,\Omega_\kappa}$ & Clarke's statistic of test to determine an equivalence in \\ & \hspace{2em} distribution between $\delta_{\ell,\Omega_\kappa}^{(1:300)}$ and $\delta_{\ell,\Omega_\kappa}^{(301:600)}$  \\
	$\bar{\alpha}$ & significance level \\
	
		  \end{tabular}}
\end{center}
\end{table}

\newpage
\section{Figures for the threshold experimental setting - illustration of the power of the test}\label{appendixThresh}

\begin{figure}[!htbp]
\begin{center}
\includegraphics[scale=0.60]{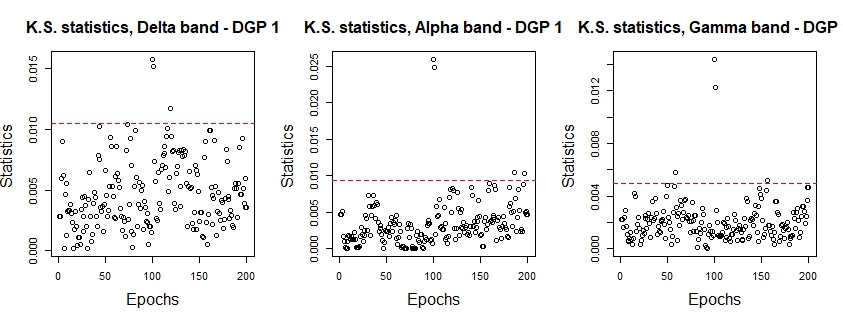}
\caption{DGP 1: Bivariate Kolmogorov-Smirnov statistics computed to compare 200 copulas, for 3 frequency bands. Red dashed line represents threshold on the Kolmogorov-Smirnov statistics for significance value of $\bar{\alpha}=1\%$.}
\label{KSdgp1}
\end{center}
\end{figure}

\newpage
\section{Tables showing the statistics of test for the four rats for Sections \ref{prepost} and \ref{twoChan} }\label{appendixPrePost}
\begin{figure}[!htbp]
{\tiny 
$$
\begin{pmatrix}
    \begin{smallmatrix} {\color{blue}1} \\ \begin{pmatrix} 118 \\ {\color{red}p=0.0444} \end{pmatrix} \end{smallmatrix} & 
		\begin{smallmatrix} {\color{blue}2} \\ \begin{pmatrix} 118 \\ {\color{red}p=0.0444} \end{pmatrix} \end{smallmatrix} &
		\begin{smallmatrix} {\color{blue}3} \\ \begin{pmatrix} 5 \\ p < 0.0001 \end{pmatrix} \end{smallmatrix} &
		\begin{smallmatrix} {\color{blue}4} \\ \begin{pmatrix} 1 \\ p< 0.0001 \end{pmatrix} \end{smallmatrix} & 
		\begin{smallmatrix} {\color{blue}5} \\ \begin{pmatrix} 49 \\ p< 0.0001 \end{pmatrix} \end{smallmatrix} &
		\begin{smallmatrix} {\color{blue}6} \\ \begin{pmatrix} 96 \\ p < 0.0001 \end{pmatrix} \end{smallmatrix} & 
		\begin{smallmatrix} {\color{blue}7} \\ \begin{pmatrix} 2 \\ p < 0.0001 \end{pmatrix} \end{smallmatrix} &
		\begin{smallmatrix} {\color{blue}8} \\ \begin{pmatrix} 2 \\ p < 0.0001 \end{pmatrix} \end{smallmatrix}
		\\ & & & & & & & \\
    \begin{smallmatrix} {\color{blue}9} \\ \begin{pmatrix} 83 \\ p< 0.0001 \end{pmatrix} \end{smallmatrix} & 
		\begin{smallmatrix} {\color{blue}10} \\ \begin{pmatrix} 45 \\ p< 0.0001 \end{pmatrix} \end{smallmatrix} &
		\begin{smallmatrix} {\color{blue}11} \\ \begin{pmatrix} 44 \\ p < 0.0001 \end{pmatrix} \end{smallmatrix} &
		\begin{smallmatrix} {\color{blue}12} \\ \begin{pmatrix} 18 \\ p< 0.0001 \end{pmatrix} \end{smallmatrix} & 
		\begin{smallmatrix} {\color{blue}13} \\ \begin{pmatrix} 134 \\ {\color{red}p=0.9515} \end{pmatrix} \end{smallmatrix} &
		\begin{smallmatrix} {\color{blue}14} \\ \begin{pmatrix} 114 \\  p=0.0125  \end{pmatrix} \end{smallmatrix} & 
		\begin{smallmatrix} {\color{blue}15} \\ \begin{pmatrix} 153 \\ {\color{red} p=0.0330 } \end{pmatrix} \end{smallmatrix} &
		\begin{smallmatrix} {\color{blue}16} \\ \begin{pmatrix} 142 \\ {\color{red} p=0.4589 } \end{pmatrix} \end{smallmatrix}
		\\ & & & & & & & \\
    \begin{smallmatrix} {\color{blue}17} \\ \begin{pmatrix} 87 \\ p < 0.0001 \end{pmatrix} \end{smallmatrix} & 
		\begin{smallmatrix} {\color{blue}18} \\ \begin{pmatrix} 62 \\ p < 0.0001 \end{pmatrix} \end{smallmatrix} &
		\begin{smallmatrix} {\color{blue}19} \\ \begin{pmatrix} 78 \\ p < 0.0001 \end{pmatrix} \end{smallmatrix} &
		\begin{smallmatrix} {\color{blue}20} \\ \begin{pmatrix} 1 \\ p < 0.0001 \end{pmatrix} \end{smallmatrix} & 
		\begin{smallmatrix} {\color{blue}21} \\ \begin{pmatrix} 60 \\ p< 0.0001 \end{pmatrix} \end{smallmatrix} &
		\begin{smallmatrix} {\color{blue}22} \\ \begin{pmatrix} 107 \\ p=0.0008 \end{pmatrix} \end{smallmatrix} & 
		\begin{smallmatrix} {\color{blue}23} \\ \begin{pmatrix} 125 \\ {\color{red} p=0.2475 } \end{pmatrix} \end{smallmatrix} &
		\begin{smallmatrix} {\color{blue}24} \\ \begin{pmatrix} 2 \\ p < 0.0001 \end{pmatrix} \end{smallmatrix}
		\\ & & & & & & &\\		
	  \begin{smallmatrix} {\color{blue}25} \\ \begin{pmatrix} 80 \\ p < 0.0001 \end{pmatrix} \end{smallmatrix} & 
		\begin{smallmatrix} {\color{blue}26} \\ \begin{pmatrix} 50 \\ p < 0.0001 \end{pmatrix} \end{smallmatrix} &
		\begin{smallmatrix} {\color{blue}27} \\ \begin{pmatrix} 81 \\ p < 0.0001 \end{pmatrix} \end{smallmatrix} &
		\begin{smallmatrix} {\color{blue}28} \\ \begin{pmatrix} 1 \\ p < 0.0001 \end{pmatrix} \end{smallmatrix} & 
		\begin{smallmatrix} {\color{blue}29} \\ \begin{pmatrix} 34 \\ p< 0.0001 \end{pmatrix} \end{smallmatrix} & 
		\begin{smallmatrix} {\color{blue}30} \\ \begin{pmatrix} 108 \\ p=0.00012 \end{pmatrix} \end{smallmatrix} & 
		\begin{smallmatrix} {\color{blue}31} \\ \begin{pmatrix} 52 \\ p < 0.0001 \end{pmatrix} \end{smallmatrix} &
		\begin{smallmatrix} {\color{blue}32} \\ \begin{pmatrix} 1 \\ p < 0.0001 \end{pmatrix} \end{smallmatrix}
		
\end{pmatrix}
$$
 }
\caption{Results of the test of difference in the equivalence pre-stroke vs post-stroke, for $\gamma$-band, displayed in the order the electrodes are places on brain. First line represents the channel index, second line the Clarke's statistic and third line the p-value related. P-values in red represent the non-rejection of $H_0$ for a significance level of $2.5 \%$, for rat id $\bf{141020}$.}
\label{algo2}
\end{figure}

\begin{figure}[!htbp]
{\tiny 
$$
\begin{pmatrix}
    \begin{smallmatrix} {\color{blue}1} \\ \begin{pmatrix} 1 \\ p < 0.0001 \end{pmatrix}  \end{smallmatrix} & 
		\begin{smallmatrix} {\color{blue}2} \\ \begin{pmatrix} 50 \\ p < 0.0001 \end{pmatrix} \end{smallmatrix} &
		\begin{smallmatrix} {\color{blue}3} \\ \begin{pmatrix} 22 \\ p < 0.0001 \end{pmatrix} \end{smallmatrix} &
		\begin{smallmatrix} {\color{blue}4} \\ \begin{pmatrix} 132 \\ {\color{red} p=0.7610} \end{pmatrix} \end{smallmatrix} & 
		\begin{smallmatrix} {\color{blue}5} \\ \begin{pmatrix} 1 \\ p< 0.0001 \end{pmatrix} \end{smallmatrix} &
		\begin{smallmatrix} {\color{blue}6} \\ \begin{pmatrix} 51 \\ p < 0.0001 \end{pmatrix} \end{smallmatrix} & 
		\begin{smallmatrix} {\color{blue}7} \\ \begin{pmatrix} 34 \\ p < 0.0001 \end{pmatrix} \end{smallmatrix} &
		\begin{smallmatrix} {\color{blue}8} \\ \begin{pmatrix} 56 \\ p < 0.0001 \end{pmatrix} \end{smallmatrix}
		\\ & & & & & & & \\
    \begin{smallmatrix} {\color{blue}9} \\ \begin{pmatrix} 1 \\ p< 0.0001 \end{pmatrix} \end{smallmatrix} & 
		\begin{smallmatrix} {\color{blue}10} \\ \begin{pmatrix} 41 \\ p< 0.0001 \end{pmatrix} \end{smallmatrix} &
		\begin{smallmatrix} {\color{blue}11} \\ \begin{pmatrix} 1 \\ p < 0.0001 \end{pmatrix} \end{smallmatrix} &
		\begin{smallmatrix} {\color{blue}12} \\ \begin{pmatrix} 10 \\ p< 0.0001 \end{pmatrix} \end{smallmatrix} & 
		\begin{smallmatrix} {\color{blue}13} \\ \begin{pmatrix} 33 \\ p< 0.0001 \end{pmatrix} \end{smallmatrix} &
		\begin{smallmatrix} {\color{blue}14} \\ \begin{pmatrix} 117 \\  {\color{red} p=0.03297}  \end{pmatrix} \end{smallmatrix} & 
		\begin{smallmatrix} {\color{blue}15} \\ \begin{pmatrix} 32 \\ p< 0.0001 \end{pmatrix} \end{smallmatrix} &
		\begin{smallmatrix} {\color{blue}16} \\ \begin{pmatrix} 5 \\ p< 0.0001 \end{pmatrix} \end{smallmatrix}
		\\ & & & & & & & \\
    \begin{smallmatrix} {\color{blue}17} \\ \begin{pmatrix} 53 \\ p < 0.0001 \end{pmatrix} \end{smallmatrix} & 
		\begin{smallmatrix} {\color{blue}18} \\ \begin{pmatrix} 13 \\ p < 0.0001 \end{pmatrix} \end{smallmatrix} &
		\begin{smallmatrix} {\color{blue}19} \\ \begin{pmatrix} 5 \\ p < 0.0001 \end{pmatrix} \end{smallmatrix} &
		\begin{smallmatrix} {\color{blue}20} \\ \begin{pmatrix} 16 \\ p < 0.0001 \end{pmatrix} \end{smallmatrix} & 
		\begin{smallmatrix} {\color{blue}21} \\ \begin{pmatrix} 25 \\ p< 0.0001 \end{pmatrix} \end{smallmatrix} &
		\begin{smallmatrix} {\color{blue}22} \\ \begin{pmatrix} 35 \\ p< 0.0001 \end{pmatrix} \end{smallmatrix} & 
		\begin{smallmatrix} {\color{blue}23} \\ \begin{pmatrix} 56 \\ p< 0.0001 \end{pmatrix} \end{smallmatrix} &
		\begin{smallmatrix} {\color{blue}24} \\ \begin{pmatrix} 28 \\ p < 0.0001 \end{pmatrix} \end{smallmatrix}
		\\ & & & & & & &\\		
	  \begin{smallmatrix} {\color{blue}25} \\ \begin{pmatrix} 20 \\ p < 0.0001 \end{pmatrix} \end{smallmatrix} & 
		\begin{smallmatrix} {\color{blue}26} \\ \begin{pmatrix} 6 \\ p < 0.0001 \end{pmatrix} \end{smallmatrix} &
		\begin{smallmatrix} {\color{blue}27} \\ \begin{pmatrix} 1 \\ p < 0.0001 \end{pmatrix} \end{smallmatrix} &
		\begin{smallmatrix} {\color{blue}28} \\ \begin{pmatrix} 1 \\ p < 0.0001 \end{pmatrix} \end{smallmatrix} & 
		\begin{smallmatrix} {\color{blue}29} \\ \begin{pmatrix} 60 \\ p< 0.0001 \end{pmatrix} \end{smallmatrix} & 
		\begin{smallmatrix} {\color{blue}30} \\ \begin{pmatrix} 36 \\ p< 0.0001 \end{pmatrix} \end{smallmatrix} & 
		\begin{smallmatrix} {\color{blue}31} \\ \begin{pmatrix} 7 \\ p < 0.0001 \end{pmatrix} \end{smallmatrix} &
		\begin{smallmatrix} {\color{blue}32} \\ \begin{pmatrix} 1 \\ p < 0.0001 \end{pmatrix} \end{smallmatrix}
		
\end{pmatrix}
$$
 }
\caption{Results of the test of difference in the equivalence pre-stroke vs post-stroke, for $\gamma$-band, displayed in the order the electrodes are places on brain. First line represents the channel index, second line the Clarke's statistic and third line the p-value related. P-values in red represent the non-rejection of $H_0$ for a significance level of $2.5 \%$, for rat id $\bf{150326}$.}
\label{algo215326}
\end{figure}

\begin{figure}[!htbp]
{\tiny 
$$
\begin{pmatrix}
    \begin{smallmatrix} {\color{blue}1} \\ \begin{pmatrix} 269 \\ p < 0.0001 \end{pmatrix}  \end{smallmatrix} & 
		\begin{smallmatrix} {\color{blue}2} \\ \begin{pmatrix} 266 \\ p < 0.0001 \end{pmatrix} \end{smallmatrix} &
		\begin{smallmatrix} {\color{blue}3} \\ \begin{pmatrix} 268 \\ p < 0.0001 \end{pmatrix} \end{smallmatrix} &
		\begin{smallmatrix} {\color{blue}4} \\ \begin{pmatrix} 267 \\ p < 0.0001 \end{pmatrix} \end{smallmatrix} & 
		\begin{smallmatrix} {\color{blue}5} \\ \begin{pmatrix} 266 \\ p < 0.0001 \end{pmatrix} \end{smallmatrix} &
		\begin{smallmatrix} {\color{blue}6} \\ \begin{pmatrix} 267 \\ p < 0.0001 \end{pmatrix} \end{smallmatrix} & 
		\begin{smallmatrix} {\color{blue}7} \\ \begin{pmatrix} 267 \\ p < 0.0001 \end{pmatrix} \end{smallmatrix} &
		\begin{smallmatrix} {\color{blue}8} \\ \begin{pmatrix} 266 \\ p < 0.0001 \end{pmatrix} \end{smallmatrix}
		\\ & & & & & & & \\
    \begin{smallmatrix} {\color{blue}9} \\ \begin{pmatrix} 268 \\ p< 0.0001 \end{pmatrix} \end{smallmatrix} & 
		\begin{smallmatrix} {\color{blue}10} \\ \begin{pmatrix} 269 \\ p< 0.0001 \end{pmatrix} \end{smallmatrix} &
		\begin{smallmatrix} {\color{blue}11} \\ \begin{pmatrix} 264 \\ p < 0.0001 \end{pmatrix} \end{smallmatrix} &
		\begin{smallmatrix} {\color{blue}12} \\ \begin{pmatrix} 267 \\ p< 0.0001 \end{pmatrix} \end{smallmatrix} & 
		\begin{smallmatrix} {\color{blue}13} \\ \begin{pmatrix} 268 \\ p< 0.0001 \end{pmatrix} \end{smallmatrix} &
		\begin{smallmatrix} {\color{blue}14} \\ \begin{pmatrix} 267 \\  p< 0.0001  \end{pmatrix} \end{smallmatrix} & 
		\begin{smallmatrix} {\color{blue}15} \\ \begin{pmatrix} 265 \\ p< 0.0001 \end{pmatrix} \end{smallmatrix} &
		\begin{smallmatrix} {\color{blue}16} \\ \begin{pmatrix} 265 \\ p< 0.0001 \end{pmatrix} \end{smallmatrix}
		\\ & & & & & & & \\
    \begin{smallmatrix} {\color{blue}17} \\ \begin{pmatrix} 265 \\ p < 0.0001 \end{pmatrix} \end{smallmatrix} & 
		\begin{smallmatrix} {\color{blue}18} \\ \begin{pmatrix} 268 \\ p < 0.0001 \end{pmatrix} \end{smallmatrix} &
		\begin{smallmatrix} {\color{blue}19} \\ \begin{pmatrix} 266 \\ p < 0.0001 \end{pmatrix} \end{smallmatrix} &
		\begin{smallmatrix} {\color{blue}20} \\ \begin{pmatrix} 269 \\ p < 0.0001 \end{pmatrix} \end{smallmatrix} & 
		\begin{smallmatrix} {\color{blue}21} \\ \begin{pmatrix} 267 \\ p< 0.0001 \end{pmatrix} \end{smallmatrix} &
		\begin{smallmatrix} {\color{blue}22} \\ \begin{pmatrix} 264 \\ p< 0.0001 \end{pmatrix} \end{smallmatrix} & 
		\begin{smallmatrix} {\color{blue}23} \\ \begin{pmatrix} 263 \\ p< 0.0001 \end{pmatrix} \end{smallmatrix} &
		\begin{smallmatrix} {\color{blue}24} \\ \begin{pmatrix} 268 \\ p < 0.0001 \end{pmatrix} \end{smallmatrix}
		\\ & & & & & & &\\		
	  \begin{smallmatrix} {\color{blue}25} \\ \begin{pmatrix} 256 \\ p < 0.0001 \end{pmatrix} \end{smallmatrix} & 
		\begin{smallmatrix} {\color{blue}26} \\ \begin{pmatrix} 267 \\ p < 0.0001 \end{pmatrix} \end{smallmatrix} &
		\begin{smallmatrix} {\color{blue}27} \\ \begin{pmatrix} 261 \\ p < 0.0001 \end{pmatrix} \end{smallmatrix} &
		\begin{smallmatrix} {\color{blue}28} \\ \begin{pmatrix} 265 \\ p < 0.0001 \end{pmatrix} \end{smallmatrix} & 
		\begin{smallmatrix} {\color{blue}29} \\ \begin{pmatrix}  266 \\ p< 0.0001 \end{pmatrix} \end{smallmatrix} & 
		\begin{smallmatrix} {\color{blue}30} \\ \begin{pmatrix} 269 \\ p< 0.0001 \end{pmatrix} \end{smallmatrix} & 
		\begin{smallmatrix} {\color{blue}31} \\ \begin{pmatrix} 268 \\ p < 0.0001 \end{pmatrix} \end{smallmatrix} &
		\begin{smallmatrix} {\color{blue}32} \\ \begin{pmatrix} 270 \\ p < 0.0001 \end{pmatrix} \end{smallmatrix}
		
\end{pmatrix}
$$
 }
\caption{Results of the test of difference in the equivalence pre-stroke vs post-stroke, for $\gamma$-band, displayed in the order the electrodes are places on brain. First line represents the channel index, second line the Clarke's statistic and third line the p-value related. P-values in red represent the non-rejection of $H_0$ for a significance level of $2.5 \%$, for rat id $\bf{150410}$.}
\label{algo150410   }
\end{figure}

\begin{figure}[!htbp]
{\tiny 
$$
\begin{pmatrix}
    \begin{smallmatrix} {\color{blue}1} \\ \begin{pmatrix} 201 \\ p < 0.0001 \end{pmatrix}  \end{smallmatrix} & 
		\begin{smallmatrix} {\color{blue}2} \\ \begin{pmatrix} 177  \\ p < 0.0001 \end{pmatrix} \end{smallmatrix} &
		\begin{smallmatrix} {\color{blue}3} \\ \begin{pmatrix}  245 \\ p < 0.0001 \end{pmatrix} \end{smallmatrix} &
		\begin{smallmatrix} {\color{blue}4} \\ \begin{pmatrix} 254  \\ p < 0.0001 \end{pmatrix} \end{smallmatrix} & 
		\begin{smallmatrix} {\color{blue}5} \\ \begin{pmatrix}  249 \\ p < 0.0001 \end{pmatrix} \end{smallmatrix} &
		\begin{smallmatrix} {\color{blue}6} \\ \begin{pmatrix}  221 \\ p < 0.0001 \end{pmatrix} \end{smallmatrix} & 
		\begin{smallmatrix} {\color{blue}7} \\ \begin{pmatrix} 213  \\ p < 0.0001 \end{pmatrix} \end{smallmatrix} &
		\begin{smallmatrix} {\color{blue}8} \\ \begin{pmatrix}  222 \\ p < 0.0001 \end{pmatrix} \end{smallmatrix}
		\\ & & & & & & & \\
    \begin{smallmatrix} {\color{blue}9} \\ \begin{pmatrix}  232 \\ p< 0.0001 \end{pmatrix} \end{smallmatrix} & 
		\begin{smallmatrix} {\color{blue}10} \\ \begin{pmatrix} 145  \\ {\color{red}p=0.2475} \end{pmatrix} \end{smallmatrix} &
		\begin{smallmatrix} {\color{blue}11} \\ \begin{pmatrix} 218  \\ p < 0.0001 \end{pmatrix} \end{smallmatrix} &
		\begin{smallmatrix} {\color{blue}12} \\ \begin{pmatrix}  175 \\ p< 0.0001 \end{pmatrix} \end{smallmatrix} & 
		\begin{smallmatrix} {\color{blue}13} \\ \begin{pmatrix} 60  \\ p< 0.0001 \end{pmatrix} \end{smallmatrix} &
		\begin{smallmatrix} {\color{blue}14} \\ \begin{pmatrix} 189  \\  p< 0.0001  \end{pmatrix} \end{smallmatrix} & 
		\begin{smallmatrix} {\color{blue}15} \\ \begin{pmatrix} 200  \\ p< 0.0001 \end{pmatrix} \end{smallmatrix} &
		\begin{smallmatrix} {\color{blue}16} \\ \begin{pmatrix} 223  \\ p< 0.0001 \end{pmatrix} \end{smallmatrix}
		\\ & & & & & & & \\
    \begin{smallmatrix} {\color{blue}17} \\ \begin{pmatrix} 195 \\ p < 0.0001 \end{pmatrix} \end{smallmatrix} & 
		\begin{smallmatrix} {\color{blue}18} \\ \begin{pmatrix} 176 \\ p < 0.0001 \end{pmatrix} \end{smallmatrix} &
		\begin{smallmatrix} {\color{blue}19} \\ \begin{pmatrix} 148 \\ {\color{red}p = 0.1280} \end{pmatrix} \end{smallmatrix} &
		\begin{smallmatrix} {\color{blue}20} \\ \begin{pmatrix} 140 \\ {\color{red}p = 0.5840} \end{pmatrix} \end{smallmatrix} & 
		\begin{smallmatrix} {\color{blue}21} \\ \begin{pmatrix} 174 \\ p< 0.0001 \end{pmatrix} \end{smallmatrix} &
		\begin{smallmatrix} {\color{blue}22} \\ \begin{pmatrix} 181 \\ p< 0.0001 \end{pmatrix} \end{smallmatrix} & 
		\begin{smallmatrix} {\color{blue}23} \\ \begin{pmatrix} 126 \\ {\color{red}p = 0.3009} \end{pmatrix} \end{smallmatrix} &
		\begin{smallmatrix} {\color{blue}24} \\ \begin{pmatrix} 210 \\ p < 0.0001 \end{pmatrix} \end{smallmatrix}
		\\ & & & & & & &\\		
	  \begin{smallmatrix} {\color{blue}25} \\ \begin{pmatrix} 143 \\ {\color{red}p = 0.3613} \end{pmatrix} \end{smallmatrix} & 
		\begin{smallmatrix} {\color{blue}26} \\ \begin{pmatrix} 142 \\ {\color{red}p = 0.4289} \end{pmatrix} \end{smallmatrix} &
		\begin{smallmatrix} {\color{blue}27} \\ \begin{pmatrix} 144 \\ {\color{red}p = 0.3009} \end{pmatrix} \end{smallmatrix} &
		\begin{smallmatrix} {\color{blue}28} \\ \begin{pmatrix} 81 \\ p < 0.0001 \end{pmatrix} \end{smallmatrix} & 
		\begin{smallmatrix} {\color{blue}29} \\ \begin{pmatrix}  160 \\ p=0.0028 \end{pmatrix} \end{smallmatrix} & 
		\begin{smallmatrix} {\color{blue}30} \\ \begin{pmatrix} 163 \\ p=0.0008 \end{pmatrix} \end{smallmatrix} & 
		\begin{smallmatrix} {\color{blue}31} \\ \begin{pmatrix} 108 \\ p =0.0012 \end{pmatrix} \end{smallmatrix} &
		\begin{smallmatrix} {\color{blue}32} \\ \begin{pmatrix} 191 \\ p < 0.0001 \end{pmatrix} \end{smallmatrix}
		
\end{pmatrix}
$$
 }
\caption{Results of the test of difference in the equivalence pre-stroke vs post-stroke, for $\gamma$-band, displayed in the order the electrodes are places on brain. First line represents the channel index, second line the Clarke's statistic and third line the p-value related. P-values in red represent the non-rejection of $H_0$ for a significance level of $2.5 \%$, for rat id $\bf{16046}$.}
\label{ prepo16046  }
\end{figure}

\begin{table}[!htbp]
\begin{center}
{\tiny \begin{tabular}{|l||*{8}{c|}}

        \cline{2-9} 
				\multicolumn{1}{c|}{} & & & & & & & & \\
      \multicolumn{1}{c|}{} & Channel 1  & Channel 2  & Channel 9 & Channel 10 & Channel 17 & Channel 18  & Channel  25 & Channel 26 \\
    \hline \hline
    Channel 1  & \cellcolor{intnull} & $12$ & $270$ & $265$ & $270$ & $266$ & $216$ & $88$\\
		           & \cellcolor{intnull} & $p<0.0001$ & $p<0.0001$ & $p<0.0001$ & $p<0.0001$ & $p<0.0001$  & $p<0.0001$ &$p<0.0001$ \\ \hline
							
    Channel 2  & $12$ & \cellcolor{intnull} & $270$ & $266$& $270$ & $267$ & $269$& $259$ \\
             & $p<0.0001$ & \cellcolor{intnull} & $p<0.0001$ & $p<0.0001$ & $p<0.0001$ & $p<0.0001$ & $p<0.0001$ & $p<0.0001$ \\ \hline	
										
    Channel 9  &$270$ & $270$ & \cellcolor{intnull} & $0$ & $268$ & $0$ & $0$ & $0$\\
		           &$p<0.0001$ & $p<0.0001$ & \cellcolor{intnull} & $p<0.0001$ & $p<0.0001$ & $p<0.0001$ & $p<0.0001$ & $p<0.0001$\\ \hline
							
    Channel 10  & $265$ & $266$& $0$ & \cellcolor{intnull} & $270$ & $8$ & $7$ & $6$ \\
             & $p<0.0001$ & $p<0.0001$ & $p<0.0001$ & \cellcolor{intnull} & $p<0.0001$ & $p<0.0001$ & $p<0.0001$ & $p<0.0001$ \\ \hline	
						
    Channel 17  & $270$ & $270$ & $268$ & $270$ & \cellcolor{intnull}& $0$ &$0$ & $0$\\
		           & $p<0.0001$ & $p<0.0001$ & $p<0.0001$ & $p<0.0001$ & \cellcolor{intnull} & $p<0.0001$& $ p<0.0001$& $p<0.0001$ \\ \hline
							
    Channel 18  & $266$ & $267$ &$0$ &$8$ & $0$ & \cellcolor{intnull} & $11$ & $6$ \\
             & $p<0.0001$ & $p<0.0001$ & $p<0.0001$ & $p<0.0001$ & $p<0.0001$ & \cellcolor{intnull} & $p < 0.0001$ & $p < 0.0001$ \\ \hline	
						
    Channel 25  & $216$ & $269$ & $0$ & $7$ & $0$& $11$ & \cellcolor{intnull} & $21$\\
		           & $p<0.0001$ & $p<0.0001$ & $p<0.0001$ & $p<0.0001$ & $p<0.0001$ & $p < 0.0001$ & \cellcolor{intnull} & $p < 0.0001$ \\ \hline
							
    Channel 26  & $88$ & $259$ &$0$ & $6$ & $0$ & $6$ & $21$ & \cellcolor{intnull} \\
             & $p<0.0001$ & $p<0.0001$ & $p<0.0001$ & $p<0.0001$ & $p<0.0001$ & $p < 0.0001$ & $p < 0.0001$ & \cellcolor{intnull}\\ \hline							

		\hline
		  \end{tabular}}
	\caption{Results of the test of difference in the dependence between two channels of the first two columns, for $\gamma$-band. First line represents the Clarke's statistic and second line the p-value related, for rat id $\bf{141020}$. }
	\label{changeChannels}
\end{center}
\end{table}

\begin{table}[!htbp]
\begin{center}
{\tiny \begin{tabular}{|l||*{8}{c|}}

        \cline{2-9} 
				\multicolumn{1}{c|}{} & & & & & & & & \\
      \multicolumn{1}{c|}{} & Channel 1  & Channel 2  & Channel 9 & Channel 10 & Channel 17 & Channel 18  & Channel  25 & Channel 26 \\
    \hline \hline
    Channel 1  & \cellcolor{intnull} & $270$ & $270$ & $270$ & $270$ & $270$ & $270$ & $270$ \\
		           & \cellcolor{intnull} & $p<0.0001$ & $p<0.0001$ & $p<0.0001$ & $p<0.0001$ & $p<0.0001$  & $p<0.0001$ &$p<0.0001$ \\ \hline
							
    Channel 2  & $270$ & \cellcolor{intnull} & $17$ & $20$& $266$ & $46$ & $266$& $268$ \\
             & $p<0.0001$ & \cellcolor{intnull} & $p<0.0001$ & $p<0.0001$ & $p<0.0001$ & $p<0.0001$ & $p<0.0001$ & $p<0.0001$ \\ \hline	
										
    Channel 9  &$270$ & $17$ & \cellcolor{intnull} & $250$  & $262$ & $260$ & $265$ & $264$\\
		           &$p<0.0001$ & $p<0.0001$ & \cellcolor{intnull} & $p<0.0001$ & $p<0.0001$ & $p<0.0001$ & $p<0.0001$ & $p<0.0001$\\ \hline
							
    Channel 10  & $270$ & $20$& $250$ & \cellcolor{intnull} & $262$ & $159$ & $267$ & $267$ \\
             & $p<0.0001$ & $p<0.0001$ & $p<0.0001$ & \cellcolor{intnull} & $p<0.0001$ & $p=0.004$ & $p<0.0001$ & $p<0.0001$ \\ \hline	
						
    Channel 17  & $270$ & $266$ & $262$ & $262$ & \cellcolor{intnull}& $13$ &$76$ & $262$\\
		           & $p<0.0001$ & $p<0.0001$ & $p<0.0001$ & $p<0.0001$ & \cellcolor{intnull} & $p<0.0001$& $ p<0.0001$& $p<0.0001$ \\ \hline
							
    Channel 18  & $270$ & $46$ &$260$ &$159$ & $13$ & \cellcolor{intnull} & $264$ & $264$ \\
             & $p<0.0001$ & $p<0.0001$ & $p<0.0001$ & $p=0.004$ & $p<0.0001$ & \cellcolor{intnull} & $p < 0.0001$ & $p < 0.0001$ \\ \hline	
						
    Channel 25  & $270$ & $266$ & $265$ & $267$ & $76$& $264$ & \cellcolor{intnull} & $256$\\
		           & $p<0.0001$ & $p<0.0001$ & $p<0.0001$ & $p<0.0001$ & $p<0.0001$ & $p < 0.0001$ & \cellcolor{intnull} & $p < 0.0001$ \\ \hline
							
    Channel 26  & $270$ & $268$ &$264$ & $267$ & $262$ & $264$ & $256$ & \cellcolor{intnull} \\
             & $p<0.0001$ & $p<0.0001$ & $p<0.0001$ & $p<0.0001$ & $p<0.0001$ & $p < 0.0001$ & $p < 0.0001$ & \cellcolor{intnull}\\ \hline							

		\hline
		  \end{tabular}}
	\caption{Results of the test of difference in the dependence between two channels of the first two columns, for $\gamma$-band. First line represents the Clarke's statistic and second line the p-value related, for rat id $\bf{150326}$.}
	\label{changeChannels150326}
\end{center}
\end{table}

\begin{table}[!htbp]
\begin{center}
{\tiny \begin{tabular}{|l||*{8}{c|}}

        \cline{2-9} 
				\multicolumn{1}{c|}{} & & & & & & & & \\
      \multicolumn{1}{c|}{} & Channel 1  & Channel 2  & Channel 9 & Channel 10 & Channel 17 & Channel 18  & Channel  25 & Channel 26 \\
    \hline \hline
    Channel 1  & \cellcolor{intnull} & $ 1$ & $1 $ & $ 184$ & $251 $ & $270 $ & $247 $ & $79 $ \\
		           & \cellcolor{intnull} & $p<0.0001$ & $p<0.0001$ & $p<0.0001$ & $p<0.0001$ & $p<0.0001$  & $p<0.0001$ &$p<0.0001$ \\ \hline
							
    Channel 2  & $ 1$ & \cellcolor{intnull} & $242$ & $233$& $259 $ & $265$ & $250$& $178$ \\
             & $p<0.0001$ & \cellcolor{intnull} & $p<0.0001$ & $p<0.0001$ & $p<0.0001$ & $p<0.0001$ & $p<0.0001$ & $p<0.0001$ \\ \hline	
										
    Channel 9  &$1 $ & $242$ & \cellcolor{intnull} & $219$  & $270$ & $270$ & $268$ & $269$\\
		           &$p<0.0001$ & $p<0.0001$ & \cellcolor{intnull} & $p<0.0001$ & $p<0.0001$ & $p<0.0001$ & $p<0.0001$ & $p<0.0001$\\ \hline
							
    Channel 10  & $184 $ & $233$& $219$ & \cellcolor{intnull} & $250$ & $270$ & $232$ & $75$ \\
             & $p<0.0001$ & $p<0.0001$ & $p<0.0001$ & \cellcolor{intnull} & $p<0.0001$ & $p>0.001$ & $p<0.0001$ & $p<0.0001$ \\ \hline	
						
    Channel 17  & $251 $ & $259 $ & $270$ & $250$ & \cellcolor{intnull}& $204$ &$166$ & $12$\\
		           & $p<0.0001$ & $p<0.0001$ & $p<0.0001$ & $p<0.0001$ & \cellcolor{intnull} & $p<0.0001$& $ p=0.0002$& $p<0.0001$ \\ \hline
							
    Channel 18  & $270 $ & $265$ &$270$ &$270$ & $204$ & \cellcolor{intnull} & $96$ & $14$ \\
             & $p<0.0001$ & $p<0.0001$ & $p<0.0001$ & $p>0.001$ & $p<0.0001$ & \cellcolor{intnull} & $p < 0.0001$ & $p < 0.0001$ \\ \hline	
						
    Channel 25  & $ 247 $ & $250$ & $268$ & $232$ & $166$& $96$ & \cellcolor{intnull} & $14$\\
		           & $p<0.0001$ & $p<0.0001$ & $p<0.0001$ & $p<0.0001$ & $p=0.0002$ & $p < 0.0001$ & \cellcolor{intnull} & $p < 0.0001$ \\ \hline
							
    Channel 26  & $79 $ & $178$ &$269$ & $75$ & $12$ & $14$ & $14$ & \cellcolor{intnull} \\
             & $p<0.0001$ & $p<0.0001$ & $p<0.0001$ & $p<0.0001$ & $p<0.0001$ & $p < 0.0001$ & $p < 0.0001$ & \cellcolor{intnull}\\ \hline							

		\hline
		  \end{tabular}}
	\caption{Results of the test of difference in the dependence between two channels of the first two columns, for $\gamma$-band. First line represents the Clarke's statistic and second line the p-value related, for rat id $\bf{150410}$.}
	\label{changeChannels150410}
\end{center}
\end{table}

\begin{table}[!htbp]
\begin{center}
{\tiny \begin{tabular}{|l||*{8}{c|}}

        \cline{2-9} 
				\multicolumn{1}{c|}{} & & & & & & & & \\
      \multicolumn{1}{c|}{} & Channel 1  & Channel 2  & Channel 9 & Channel 10 & Channel 17 & Channel 18  & Channel  25 & Channel 26 \\
    \hline \hline
    Channel 1  & \cellcolor{intnull} & $ 1$ & $1 $ & $ 184$ & $251 $ & $270 $ & $247 $ & $79 $ \\
		           & \cellcolor{intnull} & $p<0.0001$ & $p<0.0001$ & $p<0.0001$ & $p<0.0001$ & $p<0.0001$  & $p<0.0001$ &$p<0.0001$ \\ \hline
							
    Channel 2  & $ 1$ & \cellcolor{intnull} & $242$ & $233$& $259 $ & $265$ & $250$& $178$ \\
             & $p<0.0001$ & \cellcolor{intnull} & $p<0.0001$ & $p<0.0001$ & $p<0.0001$ & $p<0.0001$ & $p<0.0001$ & $p<0.0001$ \\ \hline	
										
    Channel 9  &$1 $ & $242$ & \cellcolor{intnull} & $219$  & $270$ & $270$ & $268$ & $269$\\
		           &$p<0.0001$ & $p<0.0001$ & \cellcolor{intnull} & $p<0.0001$ & $p<0.0001$ & $p<0.0001$ & $p<0.0001$ & $p<0.0001$\\ \hline
							
    Channel 10  & $184 $ & $233$& $219$ & \cellcolor{intnull} & $250$ & $270$ & $232$ & $75$ \\
             & $p<0.0001$ & $p<0.0001$ & $p<0.0001$ & \cellcolor{intnull} & $p<0.0001$ & $p>0.001$ & $p<0.0001$ & $p<0.0001$ \\ \hline	
						
    Channel 17  & $251 $ & $259 $ & $270$ & $250$ & \cellcolor{intnull}& $204$ &$166$ & $12$\\
		           & $p<0.0001$ & $p<0.0001$ & $p<0.0001$ & $p<0.0001$ & \cellcolor{intnull} & $p<0.0001$& $ p=0.0002$& $p<0.0001$ \\ \hline
							
    Channel 18  & $270 $ & $265$ &$270$ &$270$ & $204$ & \cellcolor{intnull} & $96$ & $14$ \\
             & $p<0.0001$ & $p<0.0001$ & $p<0.0001$ & $p>0.001$ & $p<0.0001$ & \cellcolor{intnull} & $p < 0.0001$ & $p < 0.0001$ \\ \hline	
						
    Channel 25  & $ 247 $ & $250$ & $268$ & $232$ & $166$& $96$ & \cellcolor{intnull} & $14$\\
		           & $p<0.0001$ & $p<0.0001$ & $p<0.0001$ & $p<0.0001$ & $p=0.0002$ & $p < 0.0001$ & \cellcolor{intnull} & $p < 0.0001$ \\ \hline
							
    Channel 26  & $79 $ & $178$ &$269$ & $75$ & $12$ & $14$ & $14$ & \cellcolor{intnull} \\
             & $p<0.0001$ & $p<0.0001$ & $p<0.0001$ & $p<0.0001$ & $p<0.0001$ & $p < 0.0001$ & $p < 0.0001$ & \cellcolor{intnull}\\ \hline							

		\hline
		  \end{tabular}}
	\caption{Results of the test of difference in the dependence between two channels of the first two columns, for $\gamma$-band. First line represents the Clarke's statistic and second line the p-value related, for rat id $\bf{160406}$.}
	\label{changeChannels160406}
\end{center}
\end{table}

\newpage
\section{Illustrations of the frequency band filtering for three channels, for rat id 141020}

\begin{figure}[!htbp]
\begin{center}
\includegraphics[scale=0.65 ]{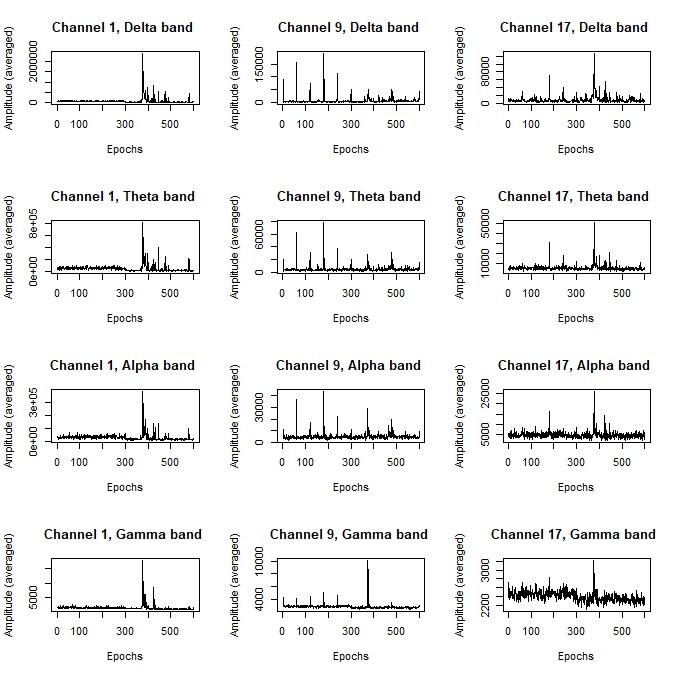}
\caption{Channels $1$, $9$ and $17$ represented through 4 of their frequency bands, for rat id $141020$.}
\label{KSdgp1}
\end{center}
\end{figure}\label{appChannels}
\newpage
\section{Proof that square root of the periodogram follows asymptotically a Rayleigh distribution}\label{appendixRay}
Let the periodogram $Z_{\ell,\Omega_\kappa}^{(r)}=(\delta_{\ell,\Omega_\kappa}^{(r)})^2$ having the asymptotic exponential distribution of density
$$g_{\ell,\Omega_\kappa}^{(r)}(Z_{\ell,\Omega_\kappa}^{(r)})=\frac{1}{\lambda}\exp \{ \frac{-Z_{\ell,\Omega_\kappa}^{(r)}}{\lambda} \} \mathds{1}_{\{Z_{\ell,\Omega_\kappa}^{(r)} > 0\}}$$
where $\lambda$ is the mean parameter. Thus, one considers the one-to-one transformation $\delta_{\ell,\Omega_\kappa}^{(r)}=\sqrt{Z_{\ell,\Omega_\kappa}^{(r)}}$. Therefore, one has the Jacobian
$$\frac{d(Z_{\ell,\Omega_\kappa}^{(r)})}{d(\delta_{\ell,\Omega_\kappa}^{(r)})}=2\delta_{\ell,\Omega_\kappa}^{(r)}.$$
Hence, the asymptotic density of $\delta_{\ell,\Omega_\kappa}^{(r)}$ is
\begin{eqnarray*}
h_{\ell,\Omega_\kappa}^{(r)}(\delta_{\ell,\Omega_\kappa}^{(r)}) &=& g_{\ell,\Omega_\kappa}^{(r)}([\delta_{\ell,\Omega_\kappa}^{(r)}]^2) \times |2\delta_{\ell,\Omega_\kappa}^{(r)}| \\
&=& \frac{2\delta_{\ell,\Omega_\kappa}^{(r)}}{\lambda} \exp \{\frac{-[\delta_{\ell,\Omega_\kappa}^{(r)}]^2}{\lambda} \}\mathds{1}_{\{\delta_{\ell,\Omega_\kappa}^{(r)} > 0\}}
\end{eqnarray*}
for $\delta_{\ell,\Omega_\kappa}^{(r)} >0$, which is the density of a Rayleigh distribution of parameter $1/\sqrt{2\lambda}$.
\end{appendices}

\end{document}